%% file: main.tex
\documentclass[twoside,openright]{report}
\usepackage{bgu-phd-thesis-prop}
\usepackage[latin1]{inputenc}
\usepackage{amsmath}
\usepackage{amsfonts}
\usepackage{amssymb}
\usepackage{graphicx}
\usepackage{epstopdf}
\usepackage{titletoc}
\usepackage{pdfpages}
\usepackage{url}
\usepackage[frozencache]{minted}
\usepackage{caption}
\usepackage[toc,page]{appendix}
\usepackage[lofdepth,lotdepth]{subfig}
\usepackage{latexsym}
\usepackage{listings}
\usepackage[algoruled,algochapter,procnumbered,dotocloa,lined,linesnumbered]{algorithm2e}

\usepackage{color}
\usepackage[hidelinks]{hyperref}
\usepackage{soul}

\newtheorem{theorem}{Theorem}[section]

\newtheorem{definition}[theorem]{Definition}

\department{Computer Science}

\DeclareMathAlphabet{\mathpzc}{OT1}{pzc}{m}{it}

\newcommand{\mr}{MapReduce}
\newcommand{\hd}{Hadoop}
\newcommand{\jtr}{JobTracker} % JobTracker
\newcommand{\ttr}{TaskTracker} % TaskTracker

\newcommand{\db}{\mathpzc{D}} % database
\newcommand{\dbp}{\mathpzc{D}^p} % probability database
\newcommand{\dbpe}{\mathcal{D}_e^p} % probability database example

\newcommand{\q}{\mathpzc{q}} % query

\newcommand{\fis}{frequent itemsets}
\newcommand{\cfis}{closed frequent itemsets}

\newcommand{\iap}{iterative-algorithms-problem}

\newcommand{\listofalgorithmes}{\tocfile{\listalgorithmcfname}{loa}}

\newcommand{\bookquote}[2]{
	\begin{center}
		\begin{quote}
			\textit{#1} 
			
			#2
		\end{quote}
	\end{center}
}

\makeatletter \newcommand \listoftodos{\section*{Todo list} \@starttoc{tdo}}
\RecustomVerbatimEnvironment{Verbatim}{BVerbatim}{}

\newcommand{\superscript}[1]{\ensuremath{^{\textrm{#1}}}}

\newcommand{\subscriptt}[1]{\ensuremath{_{\texttt{\scriptsize{#1}}}}}

\newcommand{\nosemic}{\SetEndCharOfAlgoLine{\relax}}% Drop semi-colon ;
\newcommand{\dosemic}{\SetEndCharOfAlgoLine{\string;}}% Reinstate
\newcommand{\pushline}{\Indp}% Indent
\newcommand{\popline}{\Indm\dosemic}% Undent

\begin{document}
	\includepdf[pages={1,{},2,{},3,{}}]{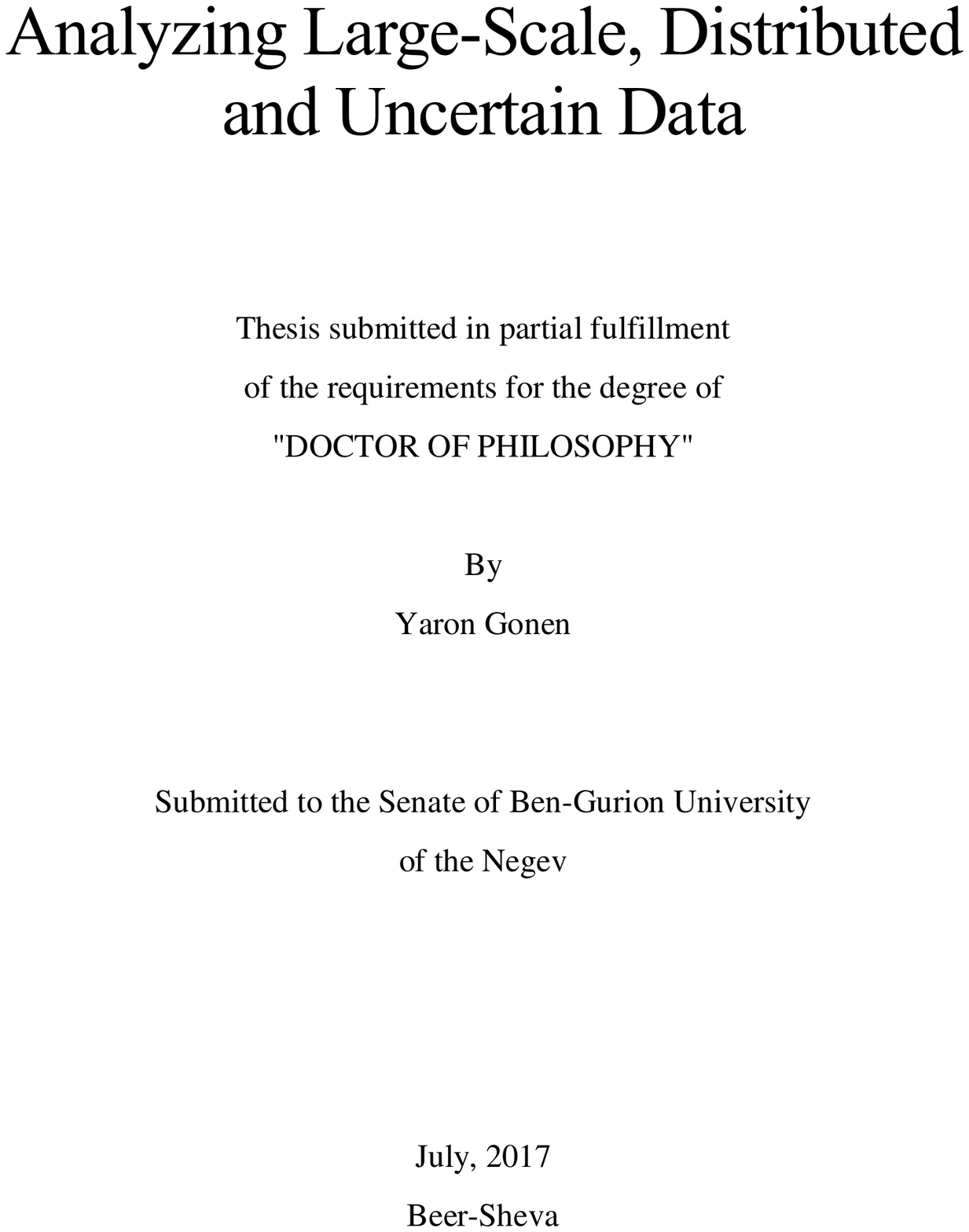}
%	\includepdf[pages={1,{}}]{../affidavit-eng.pdf}

	\title{Analyzing Large-Scale, Distributed and Uncertain Data}
	\author{Yaron Gonen}
	\advisor{Prof. Ehud Gudes}
	\copyrightfalse
	\maketitle
	\newpage
	
	\input{./general/acknowledgments}
	\newpage
	\tableofcontents
	\listoffigures
	\listofalgorithmes
	
	\pagestyle{plain}
	
	\prefacesection{Abstract}
The exponential growth of data in current times and the demand to gain \textbf{information} and \textbf{knowledge} from the data present new challenges for database researchers.
Known database systems and algorithms are no longer capable of effectively handling such large data sets.
\textit{\mr} is a novel  programming paradigm for processing distributable problems over large-scale data using a computer cluster.
The idea that lies at the core of the paradigm is to use many low-priced commodity hardware, rather than few high-end, high-priced computers and storage systems.
The key feature of \mr~ is its simplicity: the programmer needs to only implement two functions: \textit{map} and \textit{reduce}, and all the rest is handled automatically.

In this work we explore the \mr~paradigm from three different angles.

We begin by examining a well-known problem in the field of data mining: mining \mbox{\cfis} over a large dataset.
By harnessing the power of \mr, we present a novel algorithm for mining \cfis~that outperforms existing algorithms.

Next, we explore one of the fundamental implications of "Big Data": The data is not known with complete certainty.
A \textit{probabilistic} database is a relational database with the addendum that each tuple is associated with a probability of its existence.
A natural development of \mr~is of a distributed relational database management system, where relational calculus has been reduced to a combination of \textit{map} and \textit{reduce} functions.
We take this development a step further by proposing a query optimizer over distributed, probabilistic database.

Finally, we analyze the best known implementation of \mr~called \hd, aiming to overcome one of its major drawbacks.
\hd~is designed for data processing in a single pass with one \mr~job.
It does not directly support the explicit specification of the data repeatedly processed throughout different jobs.
In \hd, the input of a job is always loaded from its original location (be it a distributed file system, a database or any user-specified source).
When applying different jobs on the same data, the repeatedly processed input needs to be loaded for each job - creating a considerable disk and network I/O performance overhead.
Many data-mining algorithms, such as clustering and association-rules require iterative computation: the same data are processed again and again until the computation converges or a stopping condition is satisfied.
From \hd's perspective, iterative programs are different jobs executed one-after-the-other.
We propose a modification to \hd~such that it will support efficient access to the same data in different jobs.
We have implemented it and showed a large performance increase.

	\noindent\textbf{Keywords: Data Mining, Databases}
	\newpage
	
	%\startcontent
	\pagenumbering{arabic}

	% Introduction
	\input{./general/introduction}

	\input{./general/background}
	\input{./general/map.reduce}

	% Mining Closed Itemsets
	\input{./mining.closed.itemsets/tex/cfi_introduction}
	\input{./mining.closed.itemsets/tex/cfi_related.work}
	\input{./mining.closed.itemsets/tex/cfi_problem.definition}
	\input{./mining.closed.itemsets/tex/cfi_algorithm}

	\input{./mining.closed.itemsets/tex/cfi_experiments}
	\input{./mining.closed.itemsets/tex/cfi_conclusion}

	% Probabilistic
	\input{./probabilistic/tex/prob_introduction}
	\input{./probabilistic/tex/prob_map-reduce.rel.algebra}	
		
	\input{./probabilistic/tex/prob_example}
	\input{./probabilistic/tex/prob_opt.of.safe.and.non-safe}
	\input{./probabilistic/tex/prob_safe.plan.opt}
	\input{./probabilistic/tex/prob_experiments}
	\input{./probabilistic/tex/prob_conclusion}
	
	% Cache
	\input{./hadoop.cache/tex/cache_intro}

	\input{./hadoop.cache/tex/cache_related.work}
	\input{./hadoop.cache/tex/cache_design}

	\input{./hadoop.cache/tex/cache_experiments}
	\input{./hadoop.cache/tex/cache_conclusion}

	\input{./general/future}

	% Appendix
	\begin{appendices}
	\input{./general/apx-free-big-data-sources}
	\end{appendices}

	\bibliographystyle{plain}
	\bibliography{./bib/frequent.itemsets,./bib/prob.db,./bib/map.reduce,./bib/general}
	
%	\includepdf[pages={{},2,1}]{../heb-abstract.pdf}
%	\includepdf[pages={{},1}]{../affidavit-heb.pdf}
%s	\includepdf[pages={{},3,{},2,{},1}]{../heb-front.pdf}
\end{document}

%% file: general/acknowledgments.tex
\noindent\textbf{\Large{Acknowledgments}}\\
I am grateful to the people who made this thesis possible.
First I would like to express my deepest gratitude to my advisor, Prof. Ehud Gudes, for being there for me all the way through with his support and attentive manner.
Apart from learning how to conduct research, I learned a lot from Prof. Gudes's extensive knowledge and wide perspective in every aspect of life.
Dear Ehud, I thank you for encouraging and guiding me, I thank you for keeping your door always open, I thank you for being such a pleasant person to work with and I thank you for believing in me.

Throughout my years of study I have gained not only knowledge, but friends.
I thank the WWH team: Amnon Meiseles, Benny Lutati, Vadim Levit and Zohar Komarovsky, for being great teammates and good friends.
I thank the staff members of the "Principles of Programming Languages", with a special thanks to Mira Balaban, for teaching me the art of teaching.
A special thanks goes to Nurit Gal-Oz, a dear friend and a brilliant researcher.
We worked together over the years and I have learned a lot from her, both on the academic side and on personal issues.

I would like to thank my beloved parents, Tova and Israel, who taught me the value of education and always tried to protect me from stress and overload.
My love and appreciation goes to my son and daughter, Rotem and Tal, who learned to understand and accept my long hours of working at home.

Last but not least, I wish to thank my partner, Ariella, for encouraging me and backing me up at home.
Thank you for listening and encouraging me.
Thank you for your love and for being everything I could ever ask for in a partner.

%% file: general/introduction.tex
\chapter{Introduction}
\label{ch:introduction}

\bookquote{
	There were 5 exabytes of information created between the dawn of civilization through 2003, but that much information is now created every 2 days.
	}
	{Eric Schmidt, CEO of Google, 2010}
	
\section{The Age of Data}
This era is the era of digital data. 
Data is everywhere.
People constantly generate data:
customers buy products at retailers and the transactions are stored, smart-phone holders film videos and upload them to various video storage providers, people take pictures, and text their friends, people update their Facebook status, tweet their thoughts, leave comments and talk-backs around the web, click on ads and unceasingly generate more and more data.
Machines as well, generate and store more and more digital data:
surveillance cameras record hundreds of thousands of video hours, the Internet-of-Things and other sensor networks produce endless amount of data, satellites collect unthinkable amount of information, cars store information about the road, the status of the engine and so on.

Consider the following: 
\footnote{All the statistics are taken from \cite{big.data.facts}}
\begin{itemize}
	\item Google Photos\cite{google.photos} service, which recently announced the one-year anniversary, has 13.7 petabytes of photos that users have uploaded and 2 trillion labels.
	
	\item Facebook hosts more than 240 billion photos, growing at 7 petabytes per month.
	\item Every second, on average, around 6,000 tweets are tweeted on Twitter, which corresponds to over 350,000 tweets sent per minute, 500 million tweets per day and around 200 billion tweets per year.
\end{itemize}

The data trends are clear:
(1) every person's data footprint, which is quite large at current times, will grow significantly in the coming years, and (2) the amount of data generated by machines will be even greater than data generated by people.
Machine logs, sensor networks, retail transactions - all of these contribute to the growing pile of data.

\section{Making Sense of It}
In parallel, the amount of data made available for the public increases every year.
Retailers, hospitals, universities and other organizations understand that publishing some of their data is to their advantage.
Success in the future will be largely dictated by the ability to extract value from other organizations' data.

Initiatives such as Public Data Sets on Amazon Web Services~\cite{amazon:aws.public.data.sets} and Kaggle~\cite{kaggle} exist to accommodate the "information commons," where data can be freely (or in the case of AWS, for a modest price) shared for anyone to download and analyze.
In addition, Kaggle also hosts competitions:
it connects organizations that do not have access to advanced machine learning  with data scientists and statisticians that crave for real-world data to develop their techniques. 
A list of interesting Big Data sources can be found in Appendix~\ref{apx:list.of.big.data.repos}.

Mashups between different information sources make for unexpected and unimaginable applications that are now within reach.
For example, the Astrometry.net~\cite{lang2010astrometry} project watches the Astrometry group on Flickr~\cite{flickr} (a free photos host service) for new photos of the night sky.
It analyzes each image and identifies which part of the sky it is from, as well as any interesting celestial bodies, such as stars or galaxies.
This project shows the kind of ideas that are possible when data (such as tagged photographic images) is made available and used for something (image analysis) that was not anticipated by the creator.

\section{More Data vs. Better Algorithms}

Today, most researchers agree that \textit{"more data usually beats better algorithm"}~\cite{domingos2012few},  which is to say that for some problems (such as recommending movies or music based on past preferences), however as sophisticated as your algorithms are, they can often be beaten simply by having more data (and a less sophisticated algorithm).
On the one hand, Big Data is here, but on the other hand, we are struggling to store and analyze it.

The problem is quite simple: while the storage capacities of hard drives have increased significantly over the years, access speeds - the rate at which data can be read from drives -  have not kept up.
One typical drive from 1990 could store 1,370 MB of data and had a transfer speed of 4.4 MB/s, so you could read all the data from a full drive in around five minutes.
Over 20 years later, one terabyte drives are the norm, but the transfer speed is around 100 MB/s, so it takes more than two and a half hours to read all the data off the disk.

This is a long time to read all data on a single drive - and writing is even slower.
The obvious way to reduce the time is to read from multiple disks at once.
Imagine if we had 100 drives, each holding one hundredth of the data.
Working in parallel, we could read the data in less than two minutes.

Only using one hundredth of a disk may seem wasteful.
However, we can store one hundred datasets, each of which is one terabyte, and provide shared access to them.
We can imagine that the users of such a system would be happy to share access in return for shorter analysis times, and, statistically, that their analysis jobs would likely be spread over time, so they would not interfere with each other.

There is more to being able to read and write data in parallel to or from multiple disks, though.

The first problem to solve is a hardware failure: as soon as you start using many pieces of hardware, the chance that one will fail is fairly high.
A common way of avoiding data loss is through replication:
redundant copies of the data are kept by the system so that in the event of failure, there is another copy available.
This is how RAID works.

The second problem is that most analysis tasks need to be able to combine the data in some way; data read from one disk may need to be combined with the data from any of the other 99 disks.
Various distributed systems allow data to be combined from multiple sources, but doing this correctly is notoriously challenging.

\section{\mr}
\mr~\cite{dean2008mapreduce} provides a programming model that abstracts the problem from disk reads and writes, transforming it into a computation over sets of keys and values.
We will look at the details of this model in Chapter~\ref{ch:map.reduce}, but the important point for the present discussion is that there are two parts to the computation, the \textit{map} and the \textit{reduce}, and it is the interface between the two where the "combination" occurs.

Google was the first to publicize \mr, a system they had used to scale their data processing needs.
This system raised a lot of interest because many other businesses were facing similar scaling challenges, and it was not feasible for everyone to reinvent their own proprietary tool.
Doug Cutting, the originator of the acclaimed Lucene~\cite{lucene}, and an employee of Yahoo! at that time, saw an opportunity and led the charge to develop an open source version of this \mr~system called \hd.
Soon after, Yahoo and others rallied around to support this effort.
Today, \hd~is a core part of the computing infrastructure for many web companies, such as Yahoo, Facebook, LinkedIn, and Twitter.
Many more traditional businesses, such as media and telecom, are beginning to adopt this system as well.

\section{Thesis}
\subsection{Main Contributions}
This thesis makes the following three contributions:
\begin{itemize}
	\item A novel algorithm for mining~\cfis~over large data using the \mr~programming paradigm.
	Our experiments show that this algorithm outperforms adaptations of existing algorithms to the \mr~framework.
	A paper describing this algorithm was published in SWSTE 2016~\cite{gonen2016improved}.
	
	\item A query optimizer for relational algebra queries over a \textbf{probabilistic and distributed relational database}.
	The query execution plan, which is the output of the optimizer, is a sequence of \mr~jobs.
	We have tested our optimizer against two optimizers: one that takes into account query safety and one that does not, and our optimizer outperforms the two.
	
	\item A design and implementation of a caching mechanism for \hd, the most popular \mr~implementation to-date.
	Our experiments show that using the cache mechanism improves the performance of iterative \mr~jobs.
\end{itemize}

\subsection{Outline}

This thesis is structured as follows: Chapter \ref{ch:background} provides general background to the \mr~programming paradigm, probabilistic databases and the \cfis~ problem; 
Chapter \ref{ch:map.reduce} reviews in depth the \mr~programming paradigm: it describes its origins and components, gives examples of algorithms using it and briefly touches how the open-sourced project \hd~has implemented those components; 
In Chapter \ref{ch:mining.closed.itemsets} we introduce the \cfis~problem and present a novel algorithm for mining \cfis~using the \mr~ paradigm and the \hd~ implementation.
In Chapter \ref{ch:probabilistic} we describe the probabilistic database model with a focus on the Dalvi-Suciu Dichotomy of safe and non-safe queries, and present an optimizer of safe queries over a distributed settings using \mr.
In Chapter \ref{ch:hadoop-cache} we describe a cache mechanism for \hd~ that solves an efficiency problem that arises on iterative \mr~algorithms.
Finally, Chapter \ref{ch:future} comprises our conclusions and a discussion of future work.

%% file: general/background.tex
\chapter{Background \& Related Work}
\label{ch:background}
\bookquote{
	It is a capital mistake to theorize before one has data. Insensibly, one begins to twist the facts to suit theories, instead of theories to suit facts.
}
{A Scandal in Bohemia, Sherlock Holmes}

This chapter presents a brief background of the thesis.
A more elaborated background and related work is provided in the appropriate chapters.

\section{\mr}
\mr~\cite{dean2008mapreduce} is a programming paradigm for parallel processing of large-scale, distributed data using a computer cluster.
This paradigm was developed to answer the rising need for big data processing due to the exponential growth in stored digital data.
The idea that lies at the core of the model is to use many low-priced, commodity hardware, rather than few high-end, high-priced computers and storage systems.
The paradigm is designed in a way that relieves the programmer of the tedious job of synchronizing threads and handling processes on nodes: the programmer needs to only implement two functions: \textit{map} and \textit{reduce}, and the framework takes care of all the rest.
The most popular implementation of the \mr~model is Hadoop~\cite{hadoop}, developed by Yahoo! Labs and now maintained by Apache.

Since \mr~is one of the corner-stones of this work, a much deeper explanation of the paradigm is presented in Chapter~\ref{ch:map.reduce}.

\subsection{\hd~Architecture}

In a word, \hd~ is comprised of two components:
(a) a distributed file system called HDFS, and
(b) a framework for executing distributed \mr~ jobs.
\hd~ is deployed over a computer cluster such that each node (computer) in the cluster takes part in both of the roles above.
The two components complement one another as the file system usually holds the input for the processing job and therefore the system tries to assign the node holding the data to perform the actual processing.

A deeper coverage is presented in Chapter \ref{ch:hadoop-cache}.

\section{Probabilistic Database}
Every relational database has a hidden assumption: if a tuple is in a database, then the existence of this tuple is certain.
However, most real databases contain data which their correctness is uncertain.
In order to work with such data, there is a need to quantify the integrity of the data.
This is achieved by using \textit{probabilistic databases}.

A probabilistic database is an uncertain relational database in which the possible worlds have associated probabilities.
Probabilistic database management systems are currently an active area of research. 

Probabilistic databases distinguish between the logical data model and the physical representation of the data much like relational databases do in the ANSI-SPARC~\cite{jardine1977ansi} Architecture.
In probabilistic databases this is even more crucial since such databases have to represent very large numbers of possible worlds, often exponential in the size of one world (a classical database), succinctly.

In Chapter~\ref{ch:probabilistic} an in-depth background is presented

\section{Closed Frequent Itemsets}
We are given a large set of items (e.g. stuff sold in a supermarket) and a large set of baskets, where each basket is a small subset of the items (e.g. stuff one customer buys together).
The task is easy to explain: find the set of items that appear frequently in the baskets.
How does one define \textit{frequent}? First we define the notion of  \textit{support}: 
The support (sometimes referred as \textit{support count}) of an itemset $e$ is the number of baskets that contain $e$.
Support can be an integer, representing the actual number of baskets containing the itemset, or a percentage, representing the ratio of the baskets containing the itemset out of the total number of baskets. Given a minimum support threshold $minSup$, an itemset having a support higher than $minSup$ is a \textsl{frequent itemset}.

Agrawal and Srikant introduced the first frequent patterns mining algorithm, A-priori~\cite{agrawal1994fast}, in 1994. 
This algorithm is based on the \textsl{a-priori property}, and derives its name from it.
This property states that if a pattern $e_1$ is not frequent then any pattern $e_2$ that contains $e_1$ cannot be frequent.
In other words: \textsl{All none empty subsets of a frequent itemset must also be frequent}~\cite{han2011data}.

The number of frequent itemsets can be exponential with respect to the size of the input database.
To overcome this difficulty, the concept of \textit{closed frequent itemsets} was introduced by Pasquier et al in~\cite{pasquier1999discovering}.
An itemset $e$ is closed in a dataset $D$ if there exists no proper super-itemset $e'$ such that $e'$ has the same support as $e$ in $D$. 
An itemset $e$ is a close frequent itemset in $D$ if $e$ is closed and frequent in $d$.

Let $C$ be the set of closed frequent itemset in a dataset $D$ for a given $minSup$. Suppose that for every itemset in $C$ we are given its support count. $C$, and the given support information can be used to derive all the frequent itemsets. Therefore we can say that $C$ contains the complete information regarding frequent itemsets.

In their paper~\cite{pasquier1999discovering}, Pasquier et al presented an A-priori based algorithm called A-Close.
Pei et al proposed an efficient closed itemset mining algorithm based on the frequent pattern growth method named CLOSET~\cite{pei2000closet}, and further refined as CLOSET+~\cite{wang2003closet+}.

%% file: general/map.reduce.tex
\chapter{\mr}
\label{ch:map.reduce}

\bookquote{
	"In pioneer days they used to use oxen for heavy pulling, and when one ox couldn't budge a log, they didn't try to grow a larger ox.
	We shouldn't be trying for bigger computers, but for more systems of computers."
}
{Grace Hopper, United States Navy Rear Admiral}

\section{The Problem}
The only practical approach for dealing with Big Data problems today, with the current limitations of RAM and computer power, is the "divide-and-conquer" approach.
Divide-and-Conquer is not a new idea, but rather a fundamental concept in computer science that is introduced very early in undergraduate studies.
The basic idea is to partition a large problem into smaller sub-problems, solve each of them, and then combine all the sub-solutions into a single solution for the large problem.
To the extent that the sub-problems are independent (i.e. the process of solving each sub-problem is self-contained), they can be dealt with concurrently by different worker-threads or processes, which can be extended to many machines in a computer cluster or even to many clusters.
The sub-solutions (later will be referenced as "intermediate results") from each individual worker are then combined to produce the final result.
  
The principles behind the divide-and-conquer approach are practical for a wide range of problems in many different application domains. 
However, the details of their implementations are varied and complex. 
For example, the following are just some of the issues that need to be tackled:
\begin{itemize}
	\item \textbf{Decomposition:} 
		How to decompose a problem into independent, parallel-executable, smaller tasks?
	\item \textbf{Resource Management:}
		How to keep record of free and busy resources in a computer cluster? 
	\item \textbf{Task Assignment:} 
		How to assign tasks to workers distributed across a computer cluster, ship it to the workers, and start it?
	\item \textbf{Failure Handling:}
		How to know that a task had failed? How to handle failure of workers?
		Should the task be given to another worker or give it another try on the failed one?
	\item \textbf{Job Management:}
		How to orchestrate the entire process? 
\end{itemize}
In classic parallel or distributed programming environments, the developer needs to explicitly handle many of the above issues.
In shared memory programming, the developer needs to explicitly coordinate access to shared data structures through synchronization primitives such as mutexes, to explicitly handle process synchronization through devices such as barriers, and to remain ever vigilant
for common problems such as deadlocks and race conditions.
Language extensions, like OpenMP for shared memory parallelism, or libraries implementing the Message Passing Interface (MPI) for cluster-level parallelism, provide logical abstractions that hide details of operating system synchronization and communications primitives.
However, even with these extensions, developers are still burdened to keep track of how resources are made available to workers.
Additionally, these frameworks are mostly designed to tackle processor-intensive problems and have only rudimentary support for dealing with very large amounts of input data.
When using existing parallel computing approaches for large-data computation, the programmer must devote a significant amount of attention to low-level system details, which detracts from higher-level problem solving.

\section{The Solution}
\mr~provides to the developer an abstraction layer over the low-level details mentioned before.  
Therefore, a developer can focus on the algorithms, or computations that need to be performed, as opposed to how those computations are actually carried out or how to get the data to the processes that depend on them.

However, organizing and coordinating large amounts of computation is only part of the challenge.
Big data processing requires bringing data and code together for the computation to take place.
\mr~addresses this challenge by providing a simple abstraction for the developer, transparently handling most of the details behind the scenes in a scalable, robust, and efficient manner.
Instead of moving large amounts of data around, it is far more efficient, to move the code to the data.
This is operationally realized by spreading data across the local disks of nodes in a cluster and running processes on nodes that hold the data.
The complex task of managing storage in such a processing environment is typically handled by a distributed file system that sits underneath \mr.

\section{The Origins of \mr: \textit{map} and \textit{fold}}
The origins of \mr~lie in functional programming
\footnote{
	Functional programming is a programming paradigm that treats computation as the evaluation of mathematical functions and avoids changing state and mutable data.
	Some of the renowned functional programming languages are Common Lisp, Scheme, Haskell and F\#.
	All modern programming languages support functional programming: Java (from version 8), Python and Javascript.}.
One of the key features of functional programming is the concept of high-order functions, or, in other words, functions being first-class citizens.
It means that functions are values, just like any other primitive values (such as integers, floats, booleans etc) or complex values (such as lists, class instances etc), and can be passed as arguments to other functions.
Two of the most familiar high order functions are \textit{map} and \textit{fold}.
\textit{map} takes as an argument a function \textit{f} (that takes a single argument) and a list \textit{l}, and applies \textit{f} to all the elements in \textit{l} one by one.
\textit{fold} takes as arguments a function \textit{g} (that takes two arguments), an initial value \textit{i} and a list \textit{l}.
At first, \textit{g} is applied to \textit{i} and the first item in \textit{l}, the result of this application is stored in an intermediate variable.
This intermediate variable and the next item in \textit{l} serve as the arguments to a second application of \textit{g}, the results of which are stored in the intermediate variable (overstepping the previous value).
This process repeats until all items in \textit{l} have been processed; \textit{fold} then returns the final value of the intermediate variable.
Typically, \textit{map} and \textit{fold} are used in combination.
For example, to compute the sum of squares of a list of integers, one could map a function that squares its argument (i.e., $\lambda x.x^2$) over the input list, and then fold the resulting list with the addition function (more precisely, $\lambda x y.x+y$) using an initial value of zero.

The following Scheme code demonstrates the \textit{map} function:

\begin{minted}{scheme}
	(map (lambda (x) (* x x)) (list 1 2 3 4))
\end{minted}

The output would be:

\begin{minted}{scheme}
	(1 4 9 16)
\end{minted}

The following Scheme code demonstrates the \textit{fold}
\footnote{
	In Scheme, there are two fold function: \textit{fold-right} and \textit{fold-left}.
	In \textit{fold-right} the input lists are traversed from right to left, and in \textit{fold-left} from left to right.
}
function:

\begin{minted}{scheme}
	(fold-left (lambda (x y) (+ x y)) 0 (list 1 2 3 4))
\end{minted}

The output would be:

\begin{minted}{scheme}
	10
\end{minted}

We can think of \textit{map} as a concise way to represent the transformation of elements in a dataset (as defined by the function \textit{f}). 
Keeping a similar line-of-thought, we can think of \textit{fold} as an aggregation operation, as defined by the function \textit{g}.
One immediate observation is that the application of \textit{f} to each element in a list (or more generally, to elements in a large dataset) can be parallelized in a straightforward manner, since each functional application happens independently of other elements.
In a cluster, these operations can be distributed across many different machines.

The \textit{fold} operation, on the other hand, has more technical restrictions on the input data: elements in the list must be "downloaded" before the function \textit{g} can be applied.
However, many real-world applications do not require \textit{g} to be applied to all elements of the list.
To the extent that elements in the list can be divided into groups, the \textit{fold} aggregations can also proceed in parallel.
Furthermore, for operations that are commutative and associative, significant efficiencies can be gained in the \textit{fold} operation through local aggregation and appropriate reordering.

\section{\mr~Basics}
In a nutshell, the above is a description of \mr.
As will be elaborated later, the \mr~paradigm is composed of two phases: \textbf{map} and \textbf{reduce}.
The map phase corresponds to the \textit{map} operation in functional programming, whereas the reduce phase roughly corresponds to the \textit{fold} operation in functional programming.
The \mr~execution framework orchestrates the map and reduce phases of processing large amounts of data on large clusters of commodity machines.

From a developer perspective, \mr~provides a generic platform for processing large datasets that consists of two steps.
In the first step, a user-defined computation is applied over all input records in a dataset.
These operations occur concurrently and produce intermediate output that is later aggregated by another user-defined computation.
The programmer defines these two types of computations, and the execution framework coordinates the actual processing.
Although such a two-stage processing structure may appear to be very restrictive, many interesting algorithms can be expressed quite concisely especially if one decomposes complex algorithms into a sequence of \mr~jobs.

So what is exactly \mr?
\mr~can be associated with two different but related concepts.
First, \mr~is the programming model discussed above.
Second, it may refer to the software implementation of this programming model.

The \mr~model was originally developed by Google researches Dean and Ghemawat~\cite{dean2008mapreduce} in 2004 in an internal white paper.
Since its presentation, \mr~has enjoyed widespread implementations (Dryad~\cite{isard2007dryad} by Microsoft, Disco~\cite{disco} by Nokia, MapReduce-MPI~\cite{plimpton2011mapreduce} by Sandia National Laboratories, Phoenix~\cite{ranger2007evaluating} by Stanford and many more), but the most prominent one is \hd~\cite{hadoop}.
\hd, which is written completely in Java, acts both as a distributed file system (called HDFS) and a \mr~processing framework (called simply \hd).
It was developed by Yahoo and is now an open-source Apache project.
A vibrant software community has sprung up around \hd, with significant activity in both industry and academia.

\section{Mappers and Reducers}
The data structure that stands at the base of every processing in \mr~is the \textbf{Key-value pair}.
Keys and values may be primitive types, such as integers, floating point values, strings, and raw bytes, or they may be arbitrarily complex structures (lists, tuples, associative arrays, etc.).
Programmers typically need to define their own custom data types, although a number of libraries, such as Apache Avro~\cite{avro}, simplify the task.

Since key-value pairs play a major role in the \mr~model, the input datasets need to be described as such.
For example, for a dataset of documents, the keys may be document ids, and the values may be the actual text of the documents.
For a dataset of Twitter tweets, the keys maybe a direct URL for the tweet, and the value may be a text of the tweet.
In a relational database, 

As mentioned earlier, A \mr~computation task is composed of two sub-tasks (or steps): the \textit{mapper} step and the \textit{reducer} step.
At the core of these two steps lie two user-defined functions: \textit{map} and \textit{reduce}.
The signature of the two functions is as follows:
\[
map: (k_1, v_1)  \rightarrow [(k_2,v_2),...]
\]
\[
reduce: (k_2, [v_2,...])  \rightarrow [(k_3,v_3),...]
\]
The input to a \mr~job is a collection of key-value pairs, of the type $k_1$ and $v_1$ respectively, read from files that reside on some distributed file system (it is worth to emphasis here that the concept of \textit{key} is not a database key in the sense that it can be duplicated).

The map function is applied to every input key-value pair to generate an arbitrary number of intermediate key-value pairs.
The reduce function is applied to all values associated with the same intermediate key to generate output key-value pairs.
The transition between the two steps is called the \textit{shuffle} phase.
It is performed via a master controller that receives all the key-value pair from all the mappers, sorts them by key and divides them among the reducers so all key-value pairs with the same key end up at the same reducer.
However, no ordering relationship is guaranteed for keys across different reducers.
Output key-value pairs from each reducer are written persistently back onto the distributed file system (whereas intermediate key-value pairs are transient and not preserved).
The output ends up in \textit{r} files on the distributed file system, where \textit{r} is the number of reducers.
For the most part, there is no need to merge reducer output, since the \textit{r} files often serve as input to yet another \mr~job.
(Additional details regarding the distributed file system can be found in Chapter \ref{ch:hadoop-cache})

\begin{figure}
	\begin{center}
		\includegraphics[width=4.5in]{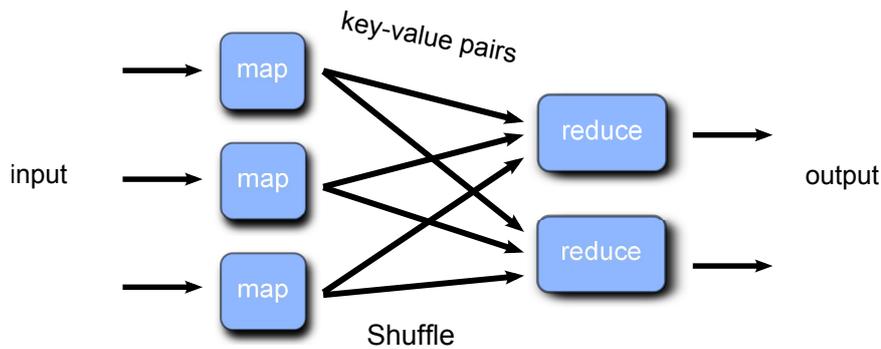}
	\end{center}
	\caption{The MapReduce data flow}
	\label{fig:map.reduce.flow}
\end{figure}

A schematic of a MapReduce computation flow can be seen in Figure~\ref{fig:map.reduce.flow}.

As an example, Algorithm~\ref{alg:words.count} is a simple word count algorithm in MapReduce.
This algorithm counts the number of occurrences of every word in a text corpus.
The input is key-value pairs in the form of $(documentId, document)$, where $documentId$ is a unique identifier of the document and $document$ is the document itself (a bag of words).
The map task gets as an input a single key-value pair of the above form, iterates over all the words in the document and for every word emits a key-value pair as follows: the word itself takes the role of the key, and the integer $1$ takes the role of the value.
The \mr~framework makes sure that all keys with the same value arrive to the same reducer.
Therefore, the reducer needs only to sum up all ones associated with each work, and emits final key-value pairs with the word as the key, and the count as the value.

\begin{algorithm}	
	\SetKwFunction{map}{map}
	\SetKwFunction{reduce}{reduce}
	\SetKwFunction{shuffle}{shuffle}
	\KwIn{$corpus$ a dataset of documents.}
	\KwResult{
		All the words that appear in any of the documents with their occurrence count.
	}
	$intermediate\_results \leftarrow \phi$\;
	\For{$doc \in corpus$}{
		$intermediate\_results.add(\map(doc))$\;
	}
	
	$group\_by \leftarrow \shuffle(intermediate\_results)$\;
	
	\For{$(word,countList) \in group\_by$}{
		\reduce($word, countList$)
	}	
	
	\caption{
		\mr~algorithm for counting the number of occurrences of each word in a documents dataset.
	}
	\label{alg:words.count}
\end{algorithm}

\begin{algorithm}	
	\SetKwFunction{emit}{emit}
	\KwIn {$doc$ The document itself (sequence of words). This is the \textit{value} part from the key-value pair.
	Specifically in this example, the key part does not play any part.
	}
	\KwResult{
		Key-Value pairs, where the key is a word from the document and the value is the number 1.
	}
	\For{$word \in doc$}{
		\emit($word$,$1$)\;
	}
	
	\caption{
		The Mapper of the word-count algorithm.
	}
	\label{alg:word-count-map}
\end{algorithm}

\begin{algorithm}	
	\SetKwFunction{emit}{emit}
	\KwIn{
		$word$ the word to be counted.
		$[c_1, c_2,...c_n]$ sequence of numbers.
		}
	\KwResult{
		A word with its count.
	}
	$sum \leftarrow 0$
	\For{$c \in [c_1, c_2,...c_n]$}{
		$sum \leftarrow sum + c$
	}
	\emit($word$, $sum$)
	\caption{The Reducer of the word-count algorithm.}
\end{algorithm}

\mr~allows for distributed processing of the map and reduce operations.
Provided each mapping operation is independent of the others, all maps can be performed in parallel - though in practice it is limited by the number of independent data sources and/or the number of CPUs near each source.
Similarly, a set of 'reducers' can perform the reduction phase - provided all outputs of the map operation that share the same key are presented to the same reducer at the same time.
While this process can often appear inefficient compared to algorithms that are more sequential, \mr~can be applied to significantly large datasets.
For example, a server farm can use \mr~to sort a petabyte of data in less than an hour~\cite{petabyteSort}.
The parallelism also offers some possibility of recovering from partial failure of servers or storage during the operation: if one mapper or reducer fails, the work can be rescheduled - assuming the input data is still available.

In addition to the "classic" \mr~processing flow, other variations are also possible.
One variation, for example, are \mr~programs that do not contain reducers, in which case mapper output is directly written to disk (one file per mapper).
This pattern fits problems that do not require aggregation, e.g., parse a large text collection or independently analyze a large number of images.
The converse is possible, with the map function being the identity function and simply passing input key-value pairs to the reducers.
This has the effect of sorting and regrouping the input for reduce-side processing.
Similarly, in some cases it is useful for the reducer to implement the identity function, in which case the program simply sorts and groups mapper output.
Finally, running identity mappers and reducers has the effect of regrouping and resorting the input data (which is sometimes useful).

\section{Partitioners and Combiners}
We have thus far presented a simplified view of \mr.
There are two additional components that complete the programming model: \textit{partitioners} and \textit{combiners}.

Partitioners are responsible for dividing up the intermediate key space and assigning intermediate key-value pairs to reducers.
In other words, the partitioner specifies the reducer task to which an intermediate key-value pair must be copied.
Within each reducer, keys are processed in sorted order.
The simplest partitioner involves computing the hash value of the key and then taking the mod of that value with the number of reducers, which can be configured. 
This method assigns approximately the same number of keys to each reducer (dependent on the quality of the hash function) and therefore creates a balance in the workloads.
Note, however, that the partitioner only considers the key and ignores the value-therefore, a roughly-even partitioning of the key space may nevertheless yield large differences in the number of key-values pairs sent to each reducer (since different keys may have different numbers of associated values).
This imbalance in the amount of data associated with each key is relatively common in many text processing applications due to the Zipfian distribution of word occurrences.

Combiners are an optimization in \mr~that allow for mapper-local aggregation before the shuffle and sort phase.
We can motivate the need for combiners by considering the word count algorithm given earlier, which emits a key-value pair for each word in the collection.
Furthermore, all these key-value pairs need to be copied across the network, and so the amount of intermediate data will be larger than the input collection itself.
This is clearly inefficient.
One solution is to perform local aggregation on the output of each mapper, i.e., to compute a local count for a word over all the documents processed by the mapper.
With this modification (assuming the maximum amount of local aggregation possible), the number of intermediate key-value pairs will be at most the number of unique words in the collection times the number of mappers (and typically far smaller because each mapper may not encounter every word).

The combiner in \mr~supports such an optimization.
One can think of combiners as "local reducers" that take place on the output of the mappers, prior to the shuffle and sort phase.
Each combiner operates in isolation and therefore does not have access to intermediate output from other mappers.
The signature of the combiner function is as follows:
\[
combine: (k_2,[v_2,...]) \rightarrow (k_2,v_2)
\]
The combiner is provided with keys and values associated with each key (the same types as the mapper output keys and values).
Critically, one cannot assume that a combiner will have the opportunity to process all values associated with the same key.
The combiner can emit any number of key-value pairs, but the keys and values must be of the same type as the mapper output (same as
the reducer input).
In cases where an operation is both associative and commutative (e.g., addition or multiplication), reducers can directly serve as combiners.
In general, however, reducers and combiners are not interchangeable.

In many cases, proper use of combiners can make the difference between an impractical algorithm and an efficient algorithm.
A combiner can significantly reduce the amount of data that needs to be copied over the network, resulting in much faster algorithms.
Output of the mappers are processed by the combiners, which perform local aggregation to cut down on the number of intermediate key-value pairs.
The partitioner determines which reducer will be responsible for processing a particular key, and the execution framework uses this
information to copy the data to the right location during the shuffle and sort phase.
Therefore, a complete \mr~job consists of code for the mapper, reducer, combiner, and partitioner, along with job configuration parameters.
The execution framework handles everything else.

\section{Communication-Cost Model}
\label{sec:communication-cost}
In general, there may be several performance measures for evaluating the performance of an algorithm in the \mr~model, such as clock time or number of flops.
In this study we follow the communication-cost model as described by Afrati and Ullman in~\cite{AfratiUllman2010}:
A \textit{task} is a single map or reduce process, executed by a single computer in the network.
The \textit{communication cost} of a task is the size of the input to this task. 
Note that the initial input to a map-task, meaning the input that resides in file, is also counted as input.
Also note that we do not distinct map-tasks from reduce-tasks for this matter.
The \textit{total communication cost} is the sum of the communications costs of all the tasks in the \mr job.

\section{Short Review of Algorithms for \mr}
We now review some common \mr~algorithms.

\subsection{Minimum, Maximum, and Count of Groups}
This algorithm calculates the minimum, maximum, and count of a field per a specific value of another field (similar to a GROUP BY in SQL).
The mapper processes input values by extracting the required fields from each input record.
The input key is ignored.
The output key is the group-by field, and the value the field we wish to aggregate.

After a grouping operation, the reducer simply iterates through all the values associated with the group and finds the minimum and maximum, as well as counts the number of members in the key grouping.

\subsection{Filtering and Joins}
An elaborated explanation regarding filtering (select) and joins appears in Chapter~\ref{ch:probabilistic}.

\subsection{Frequent Itemsets and Closed Frequent Itemsets}
One of the contributions of this thesis is a \cfis~algorithm for \mr.
This algorithm, along with a \mr~variation of the classic A-priori algorithm \cite{pasquier1999discovering} are detailed in Chapter~\ref{ch:mining.closed.itemsets}

\section{\hd~ Architecture}
As mentioned earlier, \hd~comes out-of-the-box with a distributed filesystem called HDFS, which stands for \hd~Distributed Filesystem.
HDFS is designed for data processing in a single pass: the input data is read once before the execution of the map function and is disposed at the end of the executions of all the reducers.
Therefore, a repeated read of the same data, for some iterative algorithms, creates an enormous load on the network. 

This problem is the major motivation for creating a cache mechanism for iterative \mr~algorithms, described in more details in Chapter \ref{ch:hadoop-cache}.

\section{Summary}
This chapter provides an overview of the \mr~programming model, starting with its roots in functional programming and continuing with a description of mappers, reducers, partitioners, and combiners.
Attention is also given for the performance measure defined for performance of a \mr~application.

%% file: mining.closed.itemsets/tex/cfi_introduction.tex
\chapter{Mining Closed Frequent Itemsets Using \mr}
\label{ch:mining.closed.itemsets}

\bookquote{
	Without data you're just another person with an opinion.
}{W. Edward Deming}

\section{Introduction}

Mining of \fis~for association rules has been studied thoroughly in centralized static datasets~\cite{han2007frequent} and data streams~\cite{cheng2008survey} settings.
A major branch in this research field is mining \cfis~instead of \fis~for discovering non-redundant association rules.
A set of \cfis~is proven to be a complete yet compact representation of the set of all \fis~\cite{pasquier1999discovering}.
Mining \cfis~instead of \fis~saves computation time, memory usage and produces a compacted output.
Many algorithms like Closet, Closet+, CHARM and FP-Close~\cite{han2007frequent} have been presented for mining \cfis~in centralized datasets.

Mining \cfis~in big, distributed data is more challenging than mining centralized data in the following aspects:
(1) The distributed settings is a shared-nothing environment (one can, of course, share data, however it is very expensive in terms of communication), meaning that assumptions like shared memory and shared storage, that lie at the base of most algorithms, no longer apply.
(2) Data transfer is more expensive than data processing, meaning that performance measurements change.
(3) The data is huge, and cannot reside on a single node.

MT-Closed\cite{lucchese2007parallel} and D-Closed~\cite{lucchese2010mining} are two parallel algorithms for mining \cfis, however both suffer from a few drawbacks (see section~\ref{sec:related work}) and as will be shown - our scheme overcomes these drawbacks.

\textbf{Our Contribution.}
In this chapter, we present a novel algorithm for mining \cfis~in big, distributed data settings, using the \mr~paradigm.
Using \mr~makes our algorithm very pragmatic and relatively easy to implement, maintain and execute.
In addition, our algorithm does not require a duplication elimination step, which is common to most known algorithms (it makes both the mapper and reducer more complicated, but it gives better performance).
In addition, our algorithm does not require a duplication elimination step, which is common to most known algorithms (it makes both the mapper and reducer more complicated, but it gives better performance).

%% file: mining.closed.itemsets/tex/cfi_related.work.tex
\section{Related Work}
\label{sec:related work}
The first algorithm for mining \cfis, A-Close, was introduced in~\cite{pasquier1999discovering}.
It presented the concept of \textit{generator} - a set of items that generates a single closed frequent itemset.
A-Close implements an iterative generation-and-test method for finding \cfis.
On each iteration generators are tested for frequency, and non-frequent generators are removed.
An important step is duplication elimination: generators that create an already existing itemset are also removed.
The surviving generators are used to generate the next candidate generators.
A-Close was not designed to work in a distributed setting.

\subsection{State of the Art}
MT-Closed\cite{lucchese2007parallel} is a parallel algorithm for mining \cfis.
It uses a divide-and-conquer approach on the input data to reduce the amount of data to be processed during each iteration.
However, its parallelism feature is limited.
MT-Closed is a multi-threaded algorithm designed for a multi-core architecture.
Though superior to single-core architecture, a multi-core architecture is still limited in its number of cores and its memory is limited is size and must be shared among the threads.
In addition, the input data is not distributed, and an index-building phase is required.

D-Closed~\cite{lucchese2010mining} is a shared-nothing environment, distributed algorithm for mining \cfis.
It is similar to MT-Closed in the sense that it recursively explores a sub-tree of the search space: in every iteration a candidate is generated by adding items to a previously found closure, and the dataset is projected by all the candidates.
It differs from MT-Closed in providing a clever method to detect duplicate generators: it introduces the concepts of pro-order and anti-order, and proves that among all candidates that produce the same closed itemset, only one will have no common items with its anti-order set.
However, there are a few drawbacks to D-Closed:
(1) it requires a pre-processing phase that scans the data and builds an index that needs to be shared among all the nodes, 
(2) The set of all possible items needs also to be shared among all the nodes, and
(3) the input data to each recursion call is different, meaning that iteration-wise optimizations, like caching, cannot be used.

Wang et al~\cite{wang2012mapreduce} have proposed a parallelized AFOPT-close algorithm~\cite{liu2003afopt} and have implemented it using \mr.
Like the previous algorithms, it also works in a divide-and-conquer way:
first a global list of frequent items is built, then a parallel mining of local~\cfis~is performed, and finally, non-global \cfis~are filtered out, leaving only the global \cfis.
However, they still require that final step of checking the globally \cfis~which might be very heavy, depending on the number of local results.

%\subsection{\mr~Communication Cost Model}
%\label{subsec:communication-cost}
%In general, there may be several performance measures for evaluating the performance of an algorithm in the \mr~ model, such as clock time or sum of all tasks run time.
%In this study we will follow the communication-cost model proposed by Afrati and Ullman in~\cite{AfratiUllman2010} to measure the efficiency of an algorithm in the \mr~model:
%A \textit{task} is a single map or reduce process, executed by a single computer in the network.
%The \textit{communication cost} of a task is the size of the input to this task. 
%Note that the initial input to a map-task, meaning the input that resides in file, is also counted as input.
%Also note that we do not distinct map-tasks from reduce-tasks for this matter.
%The \textit{total communication cost} is the sum of the communications costs of all the tasks in the MapReduce job.

%% file: mining.closed.itemsets/tex/cfi_problem.definition.tex
\section{Problem Definition}
\label{sec:problem definition}
Let $\mathcal{I}=\{i_1, i_2, ..., i_m\}$ be a set of \textit{items} with lexicographic order.
An \textit{itemset} $x$ is a set of items such that $x \subseteq \mathcal{I}$.
A \textit{transactional database} $\db=\{t_1,t_2,...t_n\}$, is a set of itemsets, each called a \textit{transaction}.
Each transaction in $\db$ is uniquely identified with a transaction identifier (TID), and assumed to be sorted lexicographically.
The difference between a transaction and an itemset is that as itemset is an arbitrary subset of $\mathcal{I}$ while a transaction is a subset of $\mathcal{I}$ that exists in $\db$ and identified by its id, $i$.
The \textit{support} of an itemset $x$ in $\db$, denoted $sup_{\db}(x)$, or simply $sup(x)$ when $\db$ is clear from the context, is the number of transactions in $\db$ that contain $x$ (sometimes it is the percentage of transactions).

Given a user-defined \textit{minimum support} denoted $minSup$, an itemset $x$ is called \textit{frequent} if $sup(x) \geq minSup$.

Let $\mathcal{T} \subseteq \db$ be a subset of transactions from $\db$ and let $x$ be an itemset.
We define the following two functions $f$ and $g$:
\[
	f(\mathcal{T}) = \{i \in \mathcal{I} | \forall t \in \mathcal{T}, i \in t\}
\]
\[
	g(x) = \{t \in \db | \forall i \in x, i \in t\}
\]
Function $f$ returns the intersection of all the transactions in $\mathcal{T}$, and function $g$ returns the set of all the transactions in $\db$ that contain $x$.
Notice that $g$ is antitone, meaning that for two itemsets $x_1$ and $x_2$:
$x_1 \subseteq x_2 \Rightarrow g(x_1) \supseteq g(x_2)$.
It is trivial to see that $sup(x) = |g(x)|$.
The function $h = f \circ g$ is called the \textit{Galois operator} or \textit{closure operator}.

An itemset $x$ is \textit{closed} in $\db$ iff $h(x)=x$.
It is equivalent to say that an itemset $x$ is closed in $\db$ iff no itemset that is a proper superset of $x$ has the same support in $\db$, exists.

Given a database $\db$ and a minimum support $minSup$, the \textit{mining closed frequent itemsets problem} is finding all frequent and closed itemsets in $\db$.

\subsection{Example}
Let $\mathcal{I} = \{a,b,c,d,e,f\}$, let $minSup=3$
and let $\db$ be the transaction database presented in table~\ref{tab:example.database}.
\begin{table}
	\centering
	\begin{tabular}{|c|c|}
		\hline
		TID & Transaction \\
		\hline
		$t_1$ & $\{a,c,d,e,f\}$\\
		$t_2$ & $\{a,b,e\}$\\
		$t_3$ & $\{c,e,f\}$\\
		$t_4$ & $\{a,c,d,f\}$\\
		$t_5$ & $\{c,e,f\}$\\
		\hline  
	\end{tabular}
	\caption{Example Database. TID is the transaction identifier.}
  \label{tab:example.database}
\end{table}
Consider itemset $\{c\}$.
It is a subset of transactions $t_1$, $t_3$, $t_4$ and $t_5$, meaning that $sup({c}) = 4$, which is greater than $minSup$.
However $\{c,f\}$, which is a proper superset of $\{c\}$, is also a subset of the same transactions.
$\{c\}$ is not a closed itemset since $sup(\{c,f\})=sup(\{c\})$.
The list of all \cfis~is:
$\{a\}$, 
$\{c,f\}$, 
$\{e\}$ and  
$\{c,e,f\}$.

We now present an algorithm for mining frequent closed itemsets in a distributed setting using the \mr~paradigm.

%% file: mining.closed.itemsets/tex/cfi_algorithm.tex
\section{The Algorithm}
\subsection{Overview}
\label{subsec:algorithm overview}

%% Add that each closed itemset c is a data structure that holds the generator of c.

Our algorithm is iterative, where each iteration is a \mr~job.
The inputs for iteration $i$ are
\begin{enumerate} % [label=\emph{\roman*})]
	\item $\db$, the transaction database and
	\item $C_{i-1}$ the set of the \cfis~found in the previous iteration ($C_0$, the input for the first iteration, is the empty set).
\end{enumerate}
The output of iteration $i$ is $C_i$, a set of \cfis~that have a generator of length $i$.
If $C_i \neq \phi$ then another iteration, $i+1$, is performed.
Otherwise the algorithm stops.
As mentioned earlier, each iteration is a \mr~job, comprised of a map phase and a reduce phase.
The map phase, which is equivalent to the $g$ function, emits sets of items called \textsl{closure generators} (or simply \textsl{generators}).
The reduce phase, which is equivalent to the $f$ function, finds the closure that each generator produces, and decides whether or not it should be added to $C_i$.
Each set added to $C_i$ is paired with its generator.
The generator is needed for the next iteration.

The output of the algorithm, which is the set of all \cfis, is the union of all $C_i$s.

Before the iteration begin, we have discovered that a preprocess phase that finds only the frequent items, greatly improves performance, even though another \mr~job is executed, and this data must be shared among all mapper tasks.
This \mr~job simply counts support of all items and keeps only the frequent one.

Pseudo-code of the algorithm is presented in Algorithm~\ref{alg:overview}.
We provide explanations of the important steps in the algorithm.

\subsection{Definitions}
To better understand the algorithm, we need some definitions:
% \bdefine
\begin{definition}
	\label{def:generator}
	Let $p$ be an itemset, and let $c$ be a closed itemset such that $h(p) = c$, then $p$ is called a \textbf{generator} of $c$.
\end{definition}
% \edefine

Note that a closed itemset might have more than one generator:
In the example above, both $\{c\}$ and $\{f\}$ are the generators of $\{c,f\}$.

\begin{definition}
	\label{def:map.task}
	An execution of a map function on a single transaction is called a \textbf{map task}.
\end{definition}

\begin{definition}
	\label{def:reduce.task}
	An execution of a reduce function on a specific key is called a \textbf{reduce task}.
\end{definition}

\subsection{Map Phase}
A map task in iteration $i$ gets 3 parameters as input: 
(1) a set of all the \cfis~(with their generator) found in the previous iteration, denoted $C_{i-1}$ (which is shared among all the mappers in the same iteration),
(2) a single transaction denoted $t$, and
(3) the set of all frequent items in $\db$ (again, this set is also shared among all the mappers in the same iteration and in all iterations).
(Note that in the \hd~implementation the mapper gets a set of transactions called \textit{split} and the mapper object calls the map function for each transaction in its own split only).

For each $c \in C_{i-1}$, if $c \subseteq t$ then $t$ holds the potential of finding a new \cfis, and we use it to form a new generator.
For each $item \in (t \setminus c)$ we check if $item$ is frequent.
If so, we concat $item$ to the generator of $c$ (denoted $c.generator$) thus creating $g$ (we denote that added item as $g.item$), a potential new generator for other \cfis.
The function emits a message where $g$ is the key and the tuple ($t$,1) is the value.
(The ``1'' is later summed up and used to count the support of the itemset).

Notice that $g$ is not only a generator but it is always a \textbf{minimal} generator.
Concatenating an item not in its closure guarantees to reach another minimal generator.
More precisely, it generates all minimal generators that are supersets of $g$ with one additional item and such that $t$ supports it.
Since all transactions are taken, every minimal generator with a support of at least 1 is emitted at some point.
A theorem is proved in Section~\ref{subsec:completeness}.

Pseudo code of the map function is presented in Algorithm~\ref{alg:mapper}.

\subsection{Combiner Phase}
A combiner is not a part of the \mr~programming paradigm but a \hd~implementation detail that minimizes the data transferred between map and reduce tasks.
\hd~gives the user the option of providing a \textit{combiner function} that is run on the map output, on the same machine running the mapper, and the output of the combiner function is the input for the reduce function.

In our implementation we have used a combiner, which is quite similar to the reducer, but much simpler.
The input to the combiner is a key and a collection of values: 
the key is the generator $g$ (which is an itemset) and the collection of values is a collection of tuples, composed of transactions $\mathcal{T}$, all containing $g$ and a number $s$ indicating the support of the tuple.

Since the combiner is ``local'' by nature, it has no use of the minimum support parameter, which must be applied in a global point of view.

The combiner sums the support of the input tuples, stores it in the variable $sum$ and then performs an intersection on the tuples, to get $t'$.

The combiner emits a message where $g$ is the key and the tuple ($t'$,$sum$) is the value.

Pseudo code of the combiner function is presented in Algorithm~\ref{alg:combiner}.

\subsection{Reduce Phase}
A reduce task gets as input a key, a collection of values and the minimum support.
The key is the generator $g$ (which is an itemset), the value is a collection $(t_1, s_1),...,(t_n,s_n)$ of $n$ tuples, composed of a transaction $t_i$, all containing $g$ and a number $s_i$ indicating the support of the tuple.
In addition, it gets as a parameter the user-given minimum support, $minSup$.

At first, the frequency property is checked: $sup(g) = \sum_{i=1}^{n}{s_i} \geq minSup$.
If so then an intersection of $t_1,...t_n$ is performed and a closure, denoted $c$, is produced.
If the item that was added in the map step is lexicographically greater than the first item in $c \setminus g$, then $c$ is a duplication and can be discarded.
Otherwise a new closed frequent itemset is discovered and is added $C_i$.

In other words, if the test in line~\ref{alg:reducer:dupelim} passes, then it is guaranteed that the same closure $c$ is found (and kept) in another reduce task - the one that will get $c$ from its first minimal generator in the lexicographical order.
A theorem is proved in Section~\ref{subsec:dup.elim}.

Pseudo code of the reduce function is presented in Algorithm~\ref{alg:reducer}.

In line~\ref{alg:intersection} in the algorithm we perform the $f$ function, which is actually an intersection of all the transactions in $\mathcal{T}$.
Notice that we do not need to read all of $\mathcal{T}$ and store in the RAM: $\mathcal{T}$ can be treated as a stream, reading transactions one at a time and performing the intersection.

% The main
\begin{algorithm}	
	\SetAlgoLined
	\SetKwFunction{Map}{Map}
	\SetKwFunction{Reduce}{Reduce}
	\SetKwFunction{MapReduceJob}{MapReduceJob}
	\SetKwFunction{findFrequentItems}{findFrequentItems}

	\KwIn{
		$minSup$: user-given minimum support\newline 
		$\db$: the database of transaction 
	}
	\KwOut{all closed frequent itemsets}
	\BlankLine
	
	$f\_items \leftarrow \findFrequentItems(\db,minSup)$ \;
	$C_0 \leftarrow \{\phi\} $\;
	$i \leftarrow 0$ \;
	\Repeat { $C_i = \phi$ } {
		$i++$ \;
		\nosemic $C_i \leftarrow$ \;
		\pushline\dosemic $\MapReduceJob (\db,minSup,C_{i-1},f\_items)$ \;
		\popline
	}	
	\Return{$\bigcup\limits_{j=1}^i C_j$} \;

	\caption{Main: Mine Closed Frequent Itemsets}
	\label{alg:overview}
\end{algorithm}

% map
\begin{algorithm}	
	\SetKwFunction{Concat}{concat}
	\SetKwFunction{Emit}{emit}
	
	\KwIn{
		$C_{i-1}$: The set of \cfis~with their generators, found in the previous iteration\newline 
		$t$: A single transaction from $\db$  \newline
		$f\_items$: The set of all frequent items
	}
	\KwOut{Key: potentially new generator \newline
		Value: transaction that contains the generator and its support
	}
	
	\ForEach{$c \in C_{i-1}$}{
		\If{$c \subseteq t$ } {
			$t' \leftarrow t \setminus c$\; \label{alg:mapper add item 1}
			\ForEach{$item \in t'$}{        \label{alg:mapper add item 2}
				\If{$item \in f\_items$} {
					$g \leftarrow \Concat{c.generator,item}$\;
					\Emit{$\langle g,(t,1) \rangle$}\;				
				}
			}
		}
	}
	
	\caption{Mapper}
	\label{alg:mapper}
\end{algorithm}

% combiner
\begin{algorithm}
	\SetKwFunction{Emit}{emit}
	\KwIn{
		Key: $g$, a generator \newline 
		Values: $(t_1,s_1),...,(t_n,s_n)$: a collection of $n$ tuples such that $t_i$ is a transaction and $s_i$ is its support}
	\KwOut{Key: potentially new generator \newline
		Value: transaction that contains the generator, and its support
	}
	$sum \leftarrow \sum_{i=1}^{n} s_i$ \;
	\tcc{Applying the $f$ function on $t_1,...t_n$, i.e. intersecting $t_1,...t_n$:}
	$t' \leftarrow$ $f(t_1,...t_n)$\;
	\Emit{$\langle g,(t',sum) \rangle$}\;
	
	\caption{Combiner}
	\label{alg:combiner}
\end{algorithm}

% reduce
\begin{algorithm}
	\SetKwFunction{Emit}{emit}
	\KwIn{
		Key: $g$, a generator \newline 
		Values: $(t_1,s_1),...,(t_n,s_n)$: a collection of $n$ tuples such that $t_i$ is a transaction and $s_i$ is its support \newline 
		$minSup$: The user-given minimum support}
	\KwOut{A closed frequent itemset, if found}
	$supp \leftarrow \sum_{i=1}^{n} s_i$ \;
	\If{$supp < minSup$} {
		\Return\;
	}
	\tcc{Applying the $f$ function on $t_1,...,t_n$ , i.e. intersecting $t_1,...,t_n$:}
	$c \leftarrow$ $f(t_1,...,t_n)$\; \label{alg:intersection}
	$c.generator \leftarrow g$\;
	\tcc{Duplication elimination:}
	\If{$g.item > (c \setminus g)$} { \label{alg:reducer:dupelim}
		\Return\;
	}
	\Emit{c}\;
	
	\caption{Reducer}
	\label{alg:reducer}
\end{algorithm} 

\subsection{Run Example}
\label{subsec:run.example}
Consider the example database $\db$ in Table~\ref{tab:example.database} with a minimum support of 2 transactions (0.4 in percentage).
To simulate a distributed setting we assume that each transaction $t_i$ resides on a different machine in the network (mapper node), denoted $m_i$.

\textbf{1\superscript{st} Map Phase.}
We track node $m_1$.
Its input is the transaction $t_1$ and since this is the first iteration then $C_{i-1} = C_0 = \{\phi\}$.
For each item in the input transaction we emit a message containing the item as a key and the transaction as a value.
So the messages that $m_1$ emits are the following:
$\langle \{a\}, \{a,c,d,e,f\}\rangle$, 
$\langle \{c\}, \{a,c,d,e,f\}\rangle$, \\
$\langle \{d\}, \{a,c,d,e,f\}\rangle$, 
$\langle \{e\}, \{a,c,d,e,f\}\rangle$, \\
and $\langle \{f\}, \{a,c,d,e,f\}\rangle$. \\

\textbf{1\superscript{st} Reduce Phase.}
According to the \mr~paradigm, a reducer task is assigned to every key.
We follow the reducer tasks assigned for keys $\{a\}$, $\{c\}$ and $\{f\}$, denoted $R_a$, $R_c$ and $R_f$ respectively.

First, consider $R_a$.
According to the \mr~paradigm, this reduce task receives in addition to the key $\{a\}$ all the transactions in $\db$ that contain that key: $t_1$, $t_2$ and $t_4$.
First, we must test for frequency: there are 3 transactions containing the key.
Since $minSup=2$ we pass the frequency test and can go on.
Next, we intersect all the transactions, producing the closure $\{a\}$.
The final check is whether the closure is lexicographically larger than the generator.
In our case it is not (because the generator and closure are equal), so we add $\{a\}$ to $C_1$.

Next, consider $R_c$.
This reduce task receives the key $\{c\}$, and transactions $t_1$, $t_3$, $t_4$ and $t_5$.
Since the number of messages is 4, we pass the frequency test.
The intersection of the transactions is the closure $\{c,f\}$.
Finally, $\{c\}$ is lexicographically smaller than $\{c,f\}$, so we add $\{c,f\}$ to $C_1$.

Finally, consider $R_f$.
The transactions that contain the set $\{f\}$ are $t_1$, $t_3$, $t_4$ and $t_5$.
We pass the frequency test, but the intersection is $\{c,f\}$, just like in reduce task $R_c$, so we have a duplicate result.
However, $\{f\}$ is lexicographically greater than $\{c,f\}$, so this closure is discarded.

The final set of all \cfis~found on the first iteration is:
$C_1 = \{ \{a:a\}, \{c,f:c\}, \{e:e\} \}$
(the itemset after the semicolon is the generator of this closure).

\textbf{2\superscript{nd} Map Phase.}
As before, we follow node $m_1$.
This time the set of \cfis~is not empty, so according to the algorithm, we iterate over all $c \in C_1$.
If the input transaction $t$ contains $c$, we add to $c$ all the items in $t \setminus c$, each at a time, and emit it.
So the messages that $m_1$ emits are the following: \\
$\langle \{a,c\}, \{a,c,d,e,f\}\rangle$, 
$\langle \{a,d\}, \{a,c,d,e,f\}\rangle$, \\
$\langle \{a,e\}, \{a,c,d,e,f\}\rangle$, 
$\langle \{a,f\}, \{a,c,d,e,f\}\rangle$, \\
$\langle \{c,d\}, \{a,c,d,e,f\}\rangle$, 
$\langle \{c,e\}, \{a,c,d,e,f\}\rangle$, \\
$\langle \{c,f\}, \{a,c,d,e,f\}\rangle$.
$\langle \{e,f\}, \{a,c,d,e,f\}\rangle$. \\

\textbf{2\superscript{nd} Reduce Phase.}
Consider reduce task $R_{ac}$.
According to the \mr~paradigm, this reduce task receives all the message containing the key $\{a,c\}$, which are transactions $t_1$ and $t_4$.
Since $minSup=2$ we pass the frequency test.
Next, we consider the key $\{a,c\}$ as a generator and intersect all the transactions getting the closure $\{a,c,d,f\}$.
Final check is whether the added item (c) is lexicographically larger than the closure minus the generator.
In our case it is not, so we add $\{a,c,d,f\}$ to the set of \cfis.

The full set of \cfis~is shown in table~\ref{tab:example.database.output}.

\begin{table}
	\centering
	\begin{tabular}{|c|c|c|}
		\hline
		Closed Item set & Generator & Support \\
		\hline
$\{a\}$       & $\{a\}$   & 3 \\
$\{c,f\}$     & $\{c\}$   & 4 \\
$\{e\}$       & $\{e\}$   & 4 \\
$\{a,c,d,f\}$ & $\{a,c\}$ & 2 \\
$\{a,e\}$     & $\{a,e\}$ & 2 \\
$\{c,e,f\}$   & $\{c,e\}$ & 3 \\
		\hline
	\end{tabular}
	\caption{Complete set of \cfis~ in the example database for a minimum support of 2 transactions, 
	with their generators and support.}
  \label{tab:example.database.output}
\end{table}

Next we prove the soundness and completeness of the algorithm.

\subsection{Soundness}
\label{subsec:soundness}
The mapper phase makes sure that the input to the reducer is a key which is a subset of items $p$, and a set of all transactions that contain $p$, denoted $\mathcal{T}$.
By definition $\mathcal{T} = g(p)$.
The reducer first checks that $sup(p) \geq minSup$ by checking $|\mathcal{T}| \geq minSup$ and then performs an intersection of all the transactions in $\mathcal{T}$, which by definition is the result of the function $f(\mathcal{T})$, and outputs the result.
So, by definition, all output is the result of $f \circ g$, which is a closed frequent itemset.

\subsection{Completeness}
\label{subsec:completeness}
We need to show that the algorithm outputs all the frequent closed itemsets.
Consider $c = {i_1,...i_n}$ a closed frequent itemset (that we are not sure if it was produced).
Suppose that $c$ has no proper subset that is a closed frequent itemset.
Therefore, for all items $i_j \in c$, $sup(i_j) = sup(c)$ and $g(i_j) = g(c)$.
Therefore $h(i_j) = h(c) = c$.
Since $h(i_j) = c$ then $i_j$ is a generator of $c$, and the algorithm will output $c$ at the first iteration.

Suppose that $c$ has one or more proper subsets, each is a closed frequent itemset.
We examine the largest one and denote it $l$.
We also denote its generator $g_l$, meaning that $g(l)=g(g_l)$.
Since $g$ is antitone and since $g_l \subset c$ then $g(g_l) \supset g(c)$.
What we show next is that if we add one of the items not in $l$ to $g_l$ we will generate $c$.
% YGYG removed by a suggestions of a reviewer.
% Since both $c$ and $l$ are closed, and $l \subset c$, it is obvious that $h(l) \subset h(c)$.
Consider an item $i$, such that $i \in c \setminus l$.
Let $g_l' = g_l \cup \{i\}$.
Therefore, $g(g_l') = g(g_l) \cap g(i) = g(l) \cap g(i)$.
Assume that $g(g_l') \supset g(c)$.
It implies that $l'$ is a generator of a closed itemset $h(g_l')$ that is a proper subset of $c$ in contradiction to $l$ being the largest closed subset of $c$, therefore $g(g_l') = g(c)$, meaning that $c$ will be found by the mapper by adding an item to $g_l$ (see lines~\ref{alg:mapper add item 1}-\ref{alg:mapper add item 2} in algorithm~\ref{alg:mapper}) ~$\centerdot$

\subsection{Duplication Elimination}
\label{subsec:dup.elim}
As we saw in the run example in Section~\ref{subsec:run.example}, a closed itemset can have more than a one generator, meaning that two different reduce tasks can produce the same closed itemset.
Furthermore, these two reduce tasks can be in two different iterations.
We have to identify duplicate closed itemsets and eliminate them.
The naive way to eliminate duplications is by submitting another \mr~job that sends all identical closed itemsets to the same reducer.
However, this means we need another \mr~job for that, which greatly damages performance.
Line~\ref{alg:reducer:dupelim} in algorithm~\ref{alg:reducer} takes care of that without the need for another \mr~round.
In the run example we have already seen how it works when the duplication happens on the same round.

What we would like to show is that the duplication elimination step does not "lose" any closed itemset.

We now explain the method.

Consider itemset $c=\{i_1, i_2, ..., i_n \}$ a closed, frequent itemset, and its generator $g = \{ i_{g_1}, i_{g_2}, ...,i_{g_m}\}, m<n$, such that $h(g) = c$.
According to our algorithm, $g$ was created by adding an item to a previously found closed itemset.
We denote that itemset $f$, and the added item $i_{g_j}$ such that $g = f \cup \{i_{g_j}\}$.
Suppose that $i_{g_j} > (c \setminus g)$.
In the context of the algorithm it means that $c$ would be eliminated.
We should show that $c$ can be produced from a different generator.
Consider $i_k$ the smallest item in $(c \setminus g)$.
Since $i_k \in c$ it is frequent, and since $i_k \notin g$ then surely $i_k \notin f$, meaning that the algorithm will add it to $f$, creating $g' = f \cup \{i_k\}$.
It is possible that $h(g') \subset c$, however if we keep growing $g'$ with the smallest items, we will eventually get $c$.

%% file: mining.closed.itemsets/tex/cfi_experiments.tex
\section{Experiments}
\label{sec:experiments}
We have performed several experiments in order to verify the efficiency of our algorithm and to compare it with other renowned algorithms.
\subsection{Data}
We tested our algorithm on both real and synthetic datasets.
The real dataset was downloaded from the FIMI repository~\cite{fimi-repository} and is called ``webdocs''.
It contains close to 1.7 million transactions (each transaction is a web document) with 5.3 million distinct items (each item is a word).
The maximal length of a transaction is about 71 thousand items.
The size of the dataset is 1.4 Gigabytes.
A detailed description of the ``webdocs'' dataset that also includes various statistics can be seen in~\cite{lucchese2004webdocs}.

The synthetic dataset was generated using the IBM data generator~\cite{ibmDataGen}.
We have generated six million transactions with an average of ten items per transaction, of a total of 100k items.
The total size of the input data is 600mb.

\subsection{Setup}
We run all the experiments on the Amazon Elastic \mr~\cite{amazon.emr}~infrastructure:
each run was executed on sixteen machines; each is an SSD-based instance storage for fast I/O performance with a quad core CPU and 15 GiB of memory.
All machines run \hd~version 2.6.0 with Java 8.

\subsection{Measurement}
We used communication-cost (see Section~\ref{sec:communication-cost}) as the main measurement for comparing the performance of the different algorithms.
The input records to each map task and reduce task were simply counted and summed up and the end of the execution.
This count is performed on each machine in a distributive manner.
The implementation of \hd~provides an internal input records counter that makes the counting and summing task extremely easy.
Communication-cost is an infrastructure-free measurement, meaning that it is not affected by weaker/stronger hardware or temporary network overloads, making it our measurement of choice.
However we also measured the time of execution: we run each experiment 3 times and give the average time.

\subsection{Experiments}
We have implemented the following algorithms:
($i$) a naive adaptation of Closet to \mr~, 
($ii$) the AFOPT-close adaptation to \mr, and
($iii$) our proposed algorithm.
All algorithms were implemented in Java 8, taking advantage of its new lambda expressions support.

We ran the algorithms on the two datasets with different minimum supports, and measured the communication cost and execution time for each run.

\subsection{Results}
\begin{figure}
		\centering
		\includegraphics [width=7.5cm] {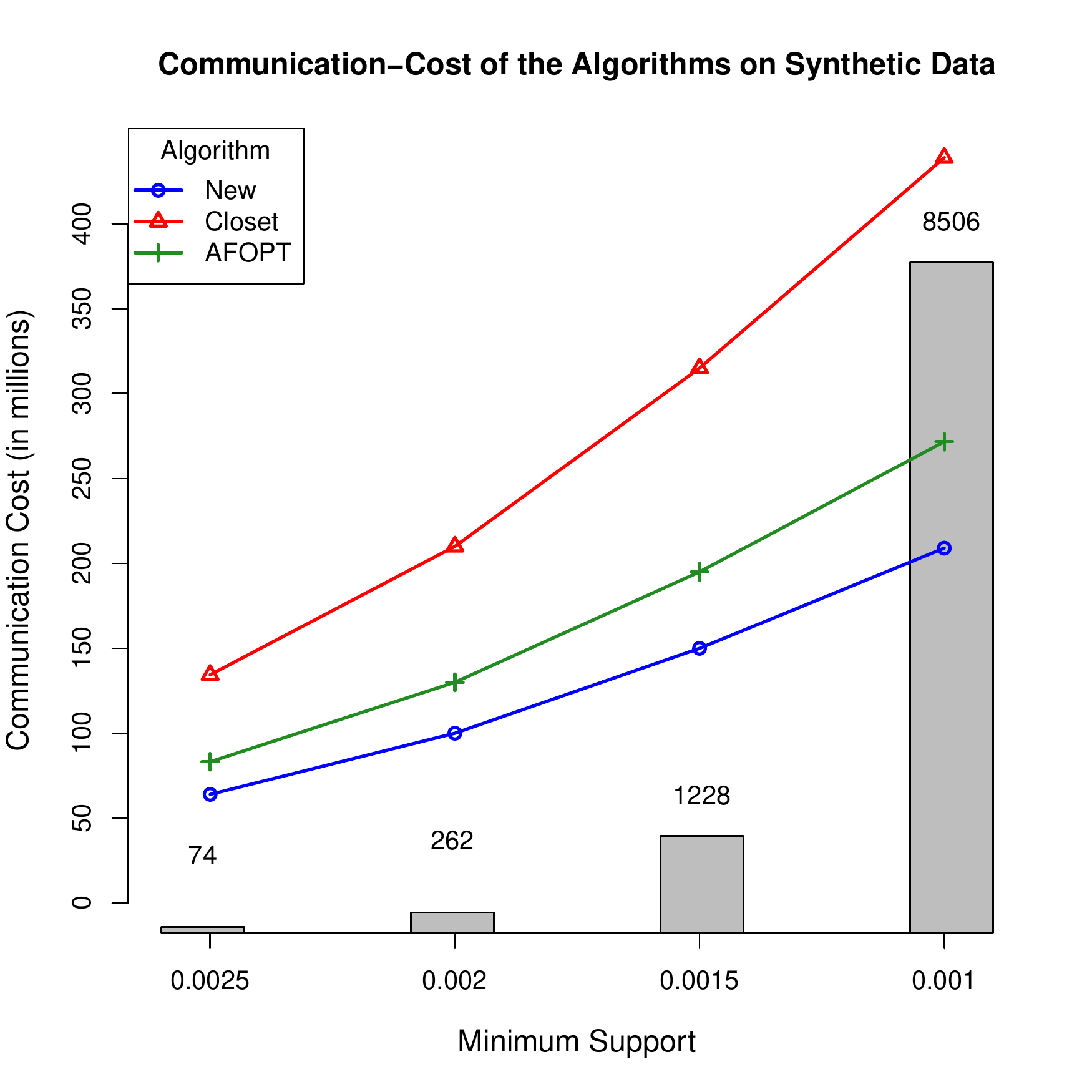}
		\caption{
			Comparing our proposed algorithm with a \mr~adaptations of Closet and AFOPT on the synthetic data.
			The lines represent the communication cost of the three algorithms for each minimum support.
			The bars present the number of \cfis~found for each minimum support.			
		}
	\label{fig:compare_alg}
\end{figure}

The first batch of runs was conducted on the synthetic dataset.
The results can be seen in figure~\ref{fig:compare_alg}:
The lines represent the communication cost of the three algorithms for each minimum support.
The bars present the number of \cfis~found for each minimum support.
As can be seen, our algorithm outperforms the others in terms of communication cost in all the minimum supports.

In the second batch of runs we run the implemented algorithms on the real dataset with four different minimum supports, and measured the communication cost and execution time for each run.

\begin{figure}
		\centering
		\includegraphics [width=7.5cm] {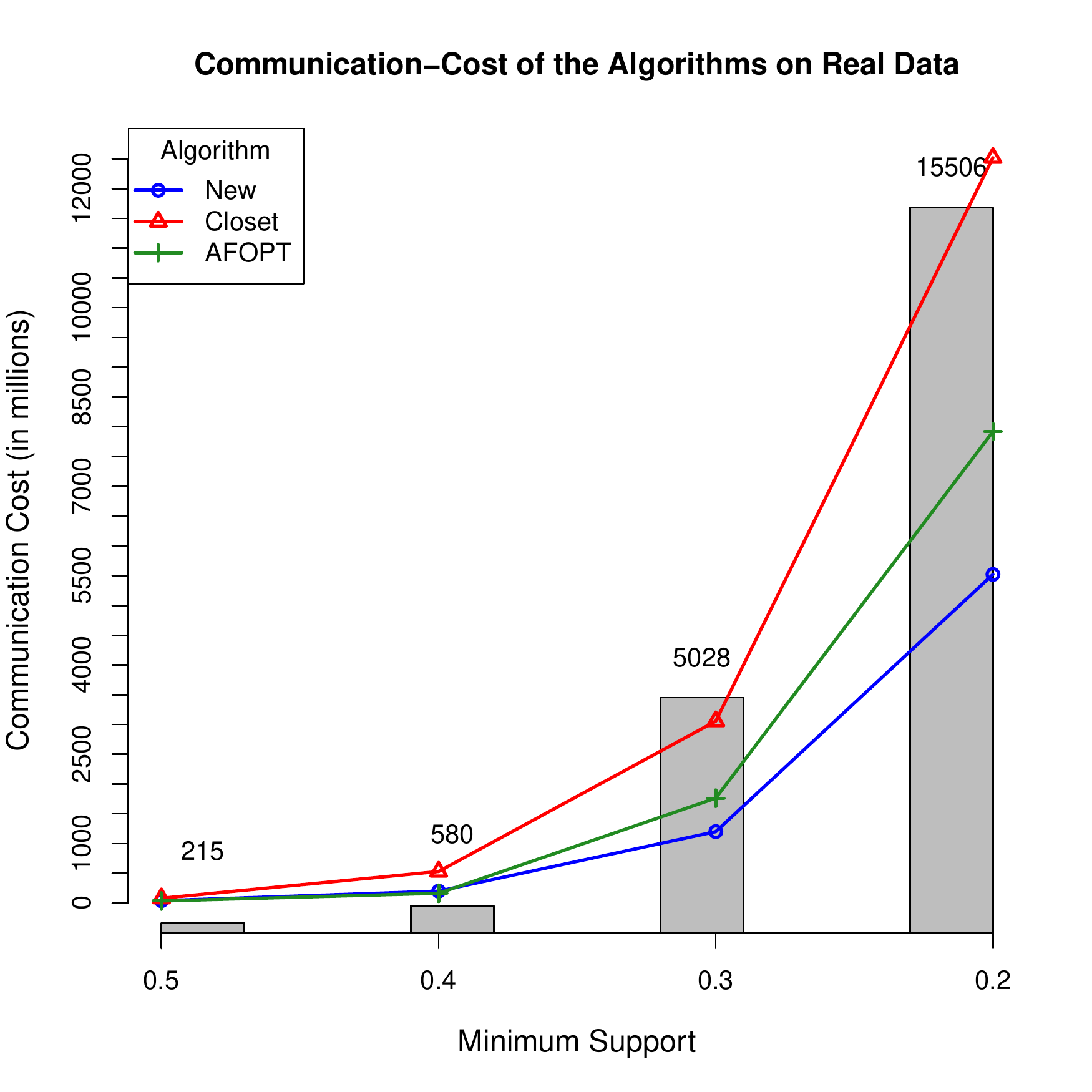}
		\caption{
			Comparing our proposed algorithm with a \mr~adaptations of Closet and AFOPT on the real data.
			The lines represent the communication cost of the three algorithms for each minimum support.
			The bars present the number of \cfis~found for each minimum support.
		}
	\label{fig:real_data}
\end{figure}

\begin{figure}
		\centering
		\includegraphics [width=7.5cm] {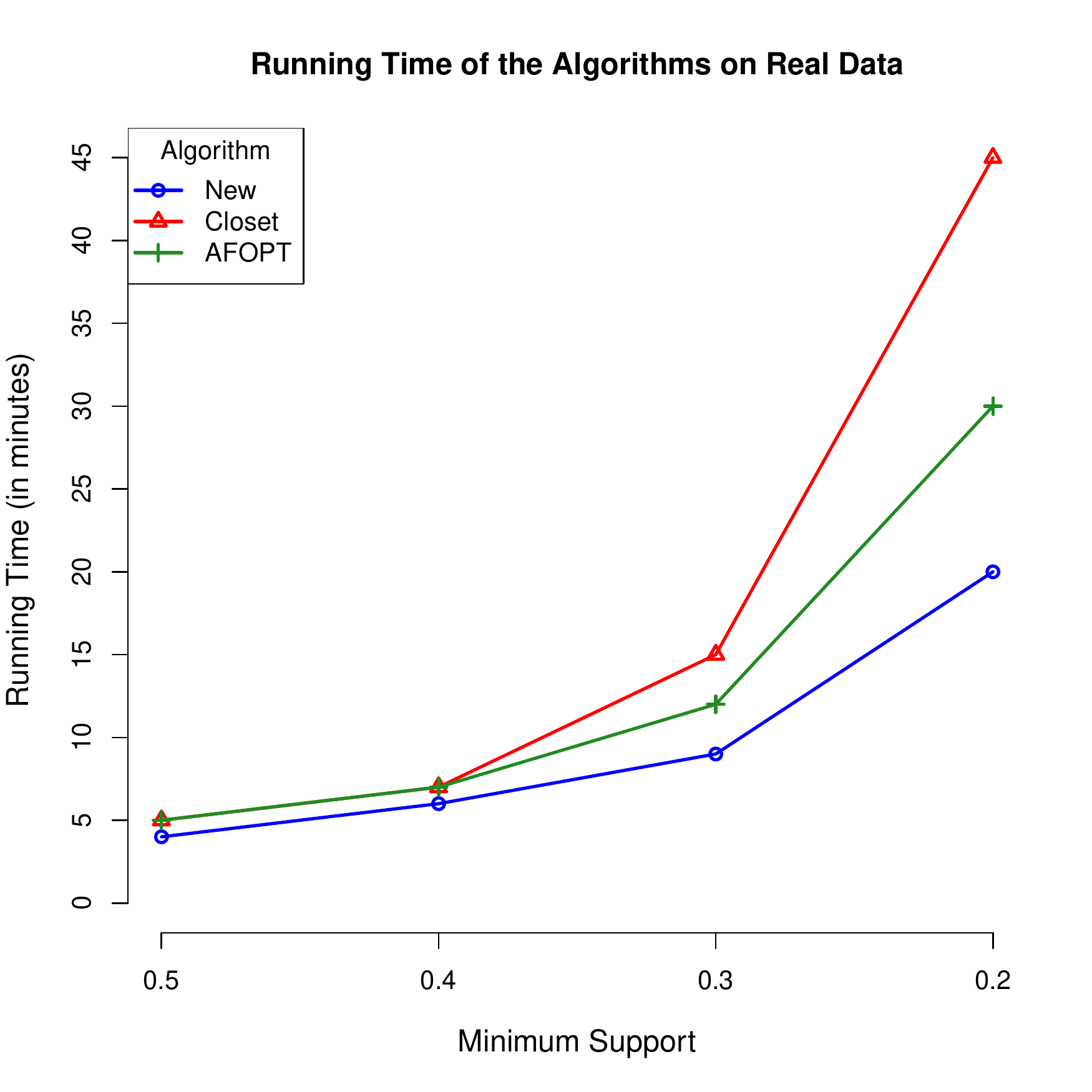}
		\caption{
			Comparing the execution time of our proposed algorithm with a \mr~adaptations of Closet and AFOPT on the real data.
		}
	\label{fig:compare_time}
\end{figure}

The results can be seen in figures~\ref{fig:real_data} and \ref{fig:compare_time}.
The lines represent the different execution times for the different minimum supports.
As can be seen, our algorithm outperforms the existing algorithms.

%% file: mining.closed.itemsets/tex/cfi_conclusion.tex
\section{Conclusion}
We have presented a new, distributed and parallel algorithm for mining \cfis~using the popular \mr~programming paradigm.
Besides its novelty, using \mr~makes this algorithm easy to implement, relieving the programmer from the wearing work of handling concurrency, synchronization and node management that are part of a distributed environment, and focus on the algorithm itself.

In addition, as mentioned in section~\ref{subsec:algorithm overview}, one of the input parameters to each iteration of the algorithm is $\db$, the database.
This parameter is the dominant one in terms of size.
Since this input is static (does not change from one iteration to the next) we might use a caching mechanism, further increasing the efficiency.
We still need to share the set of \cfis~ found in the current iteration, $C_i$, among all the nodes, however this data is significantly smaller than $\db$.

% YGYG add reference to a technical report about cache.

%% file: probabilistic/tex/prob_introduction.tex
\chapter{Query Evaluation on Distributed Probabilistic Databases Using \mr}
\label{ch:probabilistic}

\bookquote{
	In God we trust. All others must bring data.
	}{W. Edwards Deming}

\section{An Introduction to Probabilistic Databases}
\label{sec:prob.db.intro}

\subsection{Definitions}
A \textit{probabilistic database}~\cite{cavallo1987theory}, denoted $\dbp$, is a relational database, meaning that it is composed of relations, each containing tuples, with the addendum that each tuple (in all relations) is associated with a probability - a number between 0 and 1.
In this model we assume that each tuple is probabilistically independent of all other tuples.
An \textit{instance} or a \textit{possible-world} of a probabilistic database is a subset of its tuples without the probability, so actually an instance is a regular relational database.
In addition, each such instance is associated with a probability of its existence, which is calculated by the multiplications of the probabilities of its elements: the tuples that this instance contains (and of course, does not contain).
The number of possible instances of a probabilistic database $\dbp$ is $2^{\left|\dbp\right|}$ where $\left|\dbp\right|$ is the number of tuples in $\dbp$.
For a probabilistic database $\dbp$, the set of all the possible worlds is denoted $pwd(\dbp)$.
The sum of the probabilities of all possible worlds is $1.0$.

\subsection{Query Semantics on a Probabilistic Database}
On a \textbf{deterministic} relational database $\db$, a query, denoted $\q$, produces a new relation as its result.
This new relation is comprised of all the tuples in $\db$ that hold the conditions in $\q$, and is denoted $\q(\db)$

The result of a query on a \textbf{probabilistic} relational database, $\dbp$, is a new relation consisting of pairs $(t,p)$, such that $t$ is a possible tuple, meaning that it is in the query's answer in at least one of the possible worlds $pwd(\dbp)$, and $p$ is the probability that $t$ is in $\q(\mathpzc{w})$, where $\mathpzc{w}$ is chosen randomly from $pwd(\dbp)$.
In other words, $p$ represents the marginal probability of the probabilistic event "\textit{the tuple $t$ is in the result relation produced by the query $q$}" over the space of possible worlds.
In practice, $\q$ returns a set of pairs $(t_1,p_1),(t_2,p_2),...,(t_n,p_n)$, where $n$ is the number of tuples in the result relation, ordered by the probability $p_i$.

% \subsection{Notations}
% For the communication-cost calculation we use the following notations:
% \begin{itemize}
% 	\item $T(R)$ is the number of tuples of relation $R$.
% 	\item $V(R,\alpha)$ is the value count for attribute $\alpha$ of relation $R$, i.e., the number of distinct values of attribute $\alpha$ in relation $R$.
% 	Also, $V(R,[\alpha_1, \alpha_2, ..., \alpha_n ])$ is the number of distinct values of all attributes $\alpha_1, \alpha_2, ..., \alpha_n$ when considered together.
% \end{itemize}

\subsection{The Dalvi-Suciu Dichotomy}
Following the query semantics, given a query $\q$ and a probabilistic database $\dbp$, the calculation of the probability of each tuple in the result relation would require executing $\q$ over all the possible worlds (i.e. $pwd(\dbp)$), summing up the probabilities of all the possible worlds that contain this tuple.
The complexity of such procedure is linear in the number of possible worlds, which is exponential in the number of tuples in $\dbp$.

In Dalvi and Suciu's acclaimed paper~\cite{DalviSuciu:2007}, the authors show that some queries can be calculated efficiently (i.e. \textbf{not} exponential) by building an execution plan, tailored for each query, that produces the result relation with the right probability.
Such a query that can be calculated efficiently is called a \textit{safe query}, and its execution plan is called a \textit{safe plan}.
Other queries cannot have a safe plan (and are called \textit{non-safe queries}), and the only way to compute them is by iterating over all possible worlds.
This distinction between safe and non-safe queries is called the \textit{Dalvi-Suciu dichotomy}. 
In addition, for queries that have a safe plan, the authors have introduced an algorithm that produces such a plan.

\subsection{Safe-Plans, Projections and Functional Dependencies}
What makes a plan non-safe?
In simple words, an execution plan becomes non-safe when a projection operation merges two or more \textbf{probability-dependent} tuples into one.

As mentioned earlier, when two or more tuples, $t_1, ..., t_n$, are merged into a single tuple, $t_m$, the probability of $t_m$ is calculated as follows:
$p(t_m) = 1-(1-p(t_1))*...*(1-p(t_n))$.
This calculation if valid is under the assumption that all the tuples $t_1, ..., t_n$ are independent of each other.
If they are not, then this calculation is not valid and therefore the plan is not safe.

Since we assume that the tuples in all relations are independent, the case of merging dependent tuples can occur only after a join.
However, not all joins cause a dependency between tuples.
What determines if a join may cause it, is the set of functional dependencies of the database's schema.

\subsubsection{Definitions}
Consider a probabilistic database schema $\mathcal{R}$, its set of functional dependencies $\Gamma^p$ and a conjunctive query $q$.

\begin{definition}
	$Rels(q) = \{R_1,...,R_k\}, R_i \in \mathcal{R}$ is the set of all relations occurring in $q$.
\end{definition}

\begin{definition}
	Let $R$ be a probabilistic relation.
	$R.E$ is the attribute that represents the probabilistic event, or provenance.
\end{definition}

\begin{definition}
	$Attr(q)$ is the set of all attributes of all the relations in $q$.
\end{definition}

\begin{definition}
	$Head(q)$ is the set of attributes that are in the output of query $q$.
\end{definition}

\begin{definition}
	$\Gamma^p(q)$ is the \textbf{induced functional dependency} on $Attr(q)$.
	\begin{itemize}
		\item 
			Every $FD \in \Gamma^p$ is also in $\Gamma^p(q)$.
		\item 
			For every join predicate in $q$ such that $R_i.A = R_j.B$ both 
			$R_i.A \rightarrow R_j.B$ and $R_j.B \rightarrow R_i.A$
			are in $\Gamma^p(q)$.
		\item 
			For every selection predicate $R_i.A = c$, $\phi \rightarrow R_i.A$ is in 
			$\Gamma^p(q)$
	\end{itemize}
\end{definition}

%Let $q$ be a conjunctive query.
%Two relations, $R_i, R_j \in Rels(q)$ are called \textit{connected} if $q$ contains a join condition $R_i.A = R_j.B$ and either $R_i$ or $R_j$ is not in $Head(q)$.
%
%\begin{definition}
%	Given a conjunctive query $q$, relations $R_i$ and $R_j$ are called separated if they are not connected in $q$.
%\end{definition}
%
%\begin{definition}
%	Two sets of relations, $\mathcal{R}_i$, $\mathcal{R}_j$ are said to form a separation for query $q$ iff:
%	\begin{itemize}
%		\item
%			They partition $Rels(q)$.
%		\item
%			For any pair of relations $R_i$ and $R_j$ such that $R_i \in \mathcal{R}_i$ and 
%			$R_j \in \mathcal{R}_j$, they are separated.
%	\end{itemize}
%\end{definition}
%

The main theorem of safety of execution plans is the follows:

\begin{theorem}
	\label{theo:project_after_join}
	Project operator $\Pi_{A_1,...,A_k}(q)$ is \textbf{safe} iff for every 
	$R \in Rels(q)$ the following can be inferred from $\Gamma^p(q)$:
	$A_1, ..., A_k, R.E \rightarrow Head(q)$.
\end{theorem}

In simple words, Theorem \ref{theo:project_after_join} means that a projection operation is safe as long as on the result relation, no probabilistic dependent tuples will be merged.
This property is checked by examining the functional dependency of the relation to be projected:
If the projected attributes with the probabilistic event are keys in the projected table, it means one of the followings:

\begin{enumerate}
	\item The attributes themselves are not keys, but the probabilistic events are independent.
	This is usually the case when projecting a relation (i.e. not result relation of some sub query), or 
	\item The attributes are keys.
\end{enumerate}

Proof of Theorem \ref{theo:project_after_join} can be found in \cite{DalviSuciu:2007}.

As will be discussed later, this property is used in our optimization algorithm when we try to optimize project operations.

%% file: probabilistic/tex/prob_map-reduce.rel.algebra.tex
\section{Relational Algebra using \mr}
\label{sec:rel.alg.using.map.reduce}
The common relational algebra operations can be easily transferred to a combination of \textit{map} and \textit{reduce} functions, which makes \mr~applicative for a distributed relational database management system.

In this section a short review of the \mr~adaptation of the common relational algebra operators will be presented.

While most of the material is known, the presentation in this context and its communication cost analysis is new.

\subsection{Selection}
\subsubsection{Definition}
A \textit{selection} is a unary operation that specifies the tuples to retain from a relation by a condition.
A selection can have the form of $\sigma_{a\theta b}(R)$ or $\sigma_{a\theta c}(R)$, where:
\begin{itemize}
	\item $R$ is a relation
	\item $a$ and $b$ are attributes name of $R$
	\item $\theta$ is a binary operation, specifically one of the following comparison operators: $<, \le, =, \ne, \ge, >$
	\item $c$ is a constant value 
\end{itemize} 
The selection operation $\sigma_{a\theta b}(R)$ returns all the tuples in $R$ such that the condition $\theta$ holds between the attributes $a$ and $b$.
The selection operation $\sigma_{a\theta c}(R)$ returns all the tuples in $R$ such that the condition $\theta$ holds between the attributes $a$ and the constant value $c$.
The schema of the result relation is the same schema of the input relation, $R$.

More formally, the semantics of the selection operation is defined as follows:
\[\sigma_{a\theta b}(R)=\{t:t \in R, \theta(t[a],t[b]) = true\}\]
\[\sigma_{a\theta c}(R)=\{t:t \in R, \theta(t[a],c) = true\}\]

\subsubsection{\mr~Adaptation}
The selection operation can quite simply be implemented in the mapper task in the following way:
let the selection be $\sigma_{a\theta b}(R)$.
For each tuple $t$ in $R$, if $\theta(t[a], t[b])$ is true, then emit the key-value pair $(\_,t)$ (the key does not matter, since it will not be used).
If the condition does not hold, emit nothing.
Let the selection be $\sigma_{a\theta c}(R)$.
For each tuple $t$ in $R$, if $\theta(t[a], c)$ is true, than emit the key-value pair $(\_,t)$.
As before, if the condition does not hold - emit nothing.
The reducer is oblivious to this operator and can be omitted or simply be the identity function.

Pseudo-code for the \mr~adaptation of the selection operation is presented in Algorithm~\ref{alg:selection-map}.

\begin{algorithm}	
	\SetKwFunction{emit}{emit}
	\KwIn {
		$t$ a single tuples from the relation $R$.\newline
		$a,b$ attributes of $R$.\newline
		$\theta$ a binary operaion.
	}
	\KwResult{
		A key-value pair, where the key does not matter, and the value is the input tuple.
	}
	\If{$\theta(t[a], t[b])$}{
		\emit($\_$,$t$)\;
	}
	\caption{
		The Mapper of the selection operation.
	}
	\label{alg:selection-map}
\end{algorithm}

\subsubsection{Communication cost}
Each chunk of $R$ is fed to one map task, so the sum of the communication costs for all the map tasks is the size of $R$, denoted $T(R)$.
The sum of outputs of the map tasks is somewhere between $T(R)$, in the case that all tuples satisfy the condition, and $0$ in the case that none does.
So, we can say that the communication-cost of a selection is 
\[O(T(R))\]

\subsection{Projection}

\subsubsection{Definition}
A \textit{projection} is a unary operation that extracts columns from a relation.
A projection has the form of
$\Pi_{a_1,...,a_n}(R)$
where:
\begin{itemize}
	\item $a_1,...,a_n$ are attribute names
	\item $R$ is a relation
\end{itemize}
The result of such projection is the set of tuples obtained from $R$ by restricting the attributes of the tuples to $a_1,...,a_n$.
An important notion is that the result is defined as a \textbf{set}, meaning that if two tuples, $t_1$ and $t_2$, are not equal, i.e. $t_1 \ne t_2$, but the suppressed tuples are equal, i.e. 
$t_1[a_1,...,a_n] = t_2[a_1,...,a_n]$,
then the result relation will contain only one instance of the restricted tuples.

Formally, the semantics of the projection operation is defined as follows:
\[\Pi_{a_1,...,a_n}(R) = \{t[a_1,...,a_n], t \in R\}\]

\subsubsection{\mr~Adaptation}
The projection operation should be performed by the mapper because, as mentioned earlier, this operation may create duplicated tuples and the duplication-elimination process can easily be done by the reducer.
Let $\pi_A(R)$ be the projection operator, where $A=a_1,...,a_n$ is a list of attributes.
For each tuple $t$ in $R$, let $t[A]$ be a sub-tuple of $t$ that contains only the attributes in $A$. 
Emit the key-value pair $(t[A],\_)$ (the value part of the pair is of no importance, and will not be used).

The reducer function eliminates all the duplicate tuples.
In other words the reducer takes the input of $(t[A],[\_,...,\_])$, discards the list of don't-cares, then emits only the key $t[A]$.

Pseudo-codes of the mapper and the reducer are presented in Algorithms~\ref{alg:projection-map}~and~\ref{alg:projection-reduce} respectively.

\begin{algorithm}	
	\SetKwFunction{emit}{emit}
	\KwIn {
		$t$ a single tuples from the relation $R$.\newline
		$A=a_1,...,a_n$ attributes of $R$.
	}
	\KwResult{
		A key-value pair, where the key is the restricted tuples, and the value is of no importance.
	}
	\emit($t[A]$,$\_$)\;
	\caption{
		The Mapper of the projection operation.
	}
	\label{alg:projection-map}
\end{algorithm}

\begin{algorithm}	
	\SetKwFunction{emit}{emit}
	\KwIn {
		$t[A]$ the restricted tuple.
		$[\_,...,\_]$ sequence of don't-care values.
	}
	\KwResult{
		A restricted tuple.
	}
	\emit($t[A]$,$\_$)\;
	\caption{
		The Reducer of the projection operation.
	}
	\label{alg:projection-reduce}
\end{algorithm}

\subsubsection{Communication cost}
Each chunk of tuples of $R$ is fed to one map task, so the sum of the communication costs for all the map tasks is $T(R)$.
The sum of output of the map task is the same as their input.
Each output key-value pair is sent to one reduce task.
So, the communication cost of a projection is:
\[O(T(R))\]

\subsection{Natural Join}
\subsubsection{Definition}
A \textit{natural-join} is a binary operator that combines tuples from two relations that are equal on their common attribute names.
A natural-join has the form of $R \Join S$, where $R$ and $S$ are relations.
The result is a relation which is a set of all combinations of tuples in $R$ and $S$ such that the values of their common attribute names are equal.
The result is guaranteed not to have two attributes with the same name.

Formally, the semantics of the natural-join operation is defined as follows:
\[
R \Join S = \{ r \cup s | r \in R \wedge s \in S \wedge r[A] = s[A] \}
\]
Where $A$ is the set of common attributes.

\subsubsection{\mr~Adaptation}
For the natural-join task, we will look at the simple case of $R(a,b) \Join S(B,C)$.
The \textit{join} task must find all pairs of tuples from both relations that have the same value in the $B$ attribute and merge them.
In the map phase, the $B$ value will be used as the key of the key-value pair.
The value will be the complete tuple tagged with the name of the relation, so the reducer can know the original relation of each tuple.
The reducer then combines the tuples, and emits them.

For each tuple $t_r$ in $R$, the mapper emits the key-value pair $(t_r[b],(R,t_r))$, and for each tuple $t_s$ in $S$ it emits $(t_s[b],(S,t_s))$.

For all the tuples where $t_r[b] = t_s[b]$, this value is denoted $b$, and a single reducer receives the key $b$ together with the list of tagged tuples:\\
$[(R,t_{r_1}),(R,t_{r_2}),...,(R,t_{r_n}),(S,t_{s_1}),(S,t_{s_2}),...,(S,t_{s_m})]$.\\
The reducer iterates the list and combines all the tuples from tagged with $R$ with all the tuples tagged with $S$, and emits the key $b$ with the value of list of the combined tuples:\\
$(b,[(t_{r_1} \cup t_{s_1}),(t_{r_1} \cup t_{s_2}),...(t_{r_n} \cup t_{s_m})])$.

Pseudo-codes of the mapper and reducer for natural-join are presented in Algorithms~\ref{alg:join-map} and~\ref{alg:join-reduce} respectively.

\begin{algorithm}
	\SetKwFunction{Emit}{emit}
	\KwIn{
		$t$ tuple, either from $R$ or $S$
	}
	\BlankLine
	
	\eIf{$t \in R$} {
		\Emit{$t[b]$, $(R, t)$ }	\;
	}{
		\Emit{$t[b]$, $(S, t)$ }	\;
	}

	\caption{The Mapper of Natural-join}
	\label{alg:join-map}
\end{algorithm}

\begin{algorithm}
	\SetKwFunction{Emit}{emit}
	
	\KwIn{Key $b$, List$<$Tagged-Value$> [(R,t_{r_1}),..,(R,t_{r_n}),(S,t_{s_1}),..,(S,t_{s_m})]$}
	\BlankLine
	\ForAll{$(\_,t_{r_i}) \in [(R,t_{r_1}),...,(R,t_{r_n})]$}{
		\ForAll{$(\_,t_{s_j}) \in [(S,t_{s_1}),...,(S,t_{s_m})]$}{
			\Emit{$b, t_{r_i} \cup t_{s_j}$}
		}
	}
	
	\caption{The Reducer of Natural-join}
	\label{alg:join-reduce}
\end{algorithm}

\subsubsection{Communication Cost}
Each chunk of tuples of the relations $R$ and $S$ is fed into one map task, so the sum of the communications costs for all the map tasks is $T(R)+T(S)$
As before, the sum of outputs of the map tasks is roughly as large as their input.
So, the communication cost of a join is: \[O(T(R)+T(S))\]

It is worth mentioning here that Affrati and Ullman have presented another algorithm for n-way join, presented in the paper from 2010 ~\cite{AfratiUllman2010}.

\subsection{Select-Project}
\subsubsection{\mr~Adaptation}
Since a \textit{selection} operation is performed on the mapper task only, we can join it with a \textit{projection} operation easily.
Let the select-project be $\Pi_A(\sigma_{a\theta b}(R))$.
For each tuple $t$ in $R$, the mapper checks that $\theta(t[a],t[b])$ is true.
If so, let $t[A]$ be a sub-tuple of $t$ that contains only the attributes in $A$.
The mapper emits the key-value pair $(t[A],\_)$.
The reducer task acts the same as the reducer for a regular projection:
it eliminates all the duplicate restricted tuples.
In other words, the reducer turns its input $(t[A],[\_,...,\_])$ into $(t[A],t[A])$, and emits a single pair $(t[A],t[A])$.

The pseudo-code for the map and reduce functions for select-project operation are presented in Algorithms~\ref{alg:selection-projection-map}~and \ref{alg:selection-projection-reduce}~respectively. 

\begin{algorithm}
	\SetKwFunction{emit}{emit}
	\KwIn {
		$t$ a single tuples from the relation $R$.\newline
		$a,b$ attributes of $R$.\newline
		$\theta$ a binary operation.
	}
	\KwResult{
		A key-value pair, where the key does not matter, and the value is the input tuple.
	}
	\If{$\theta(t[a],t[b])$}{
		\emit($t$, $\_$)\;
	}
	\caption{The Mapper of the selection-projection operation.}	\label{alg:selection-projection-map}
\end{algorithm}

\begin{algorithm}
	\SetKwFunction{emit}{emit}
	\KwIn {
		$t[A]$ the restricted tuple.\newline
		$[\_,...,\_]$ sequence of don't-care values.
	}
	\KwResult{A restricted tuple.}
	\emit($t[A]$,$\_$)\;
	\caption{The Reducer of the selection-projection operation.}	\label{alg:selection-projection-reduce}
\end{algorithm}

\subsubsection{Communication cost}
Each chunk of $R$ is fed to one map task, so the sum of the communication costs for all the map tasks is the size of $R$, denoted $T(R)$.
The sum of outputs of the map tasks is somewhere between $T(R)$, in the case that all tuples satisfy the condition, and $0$ in the case that none does.
So, we can say that the communication-cost of a selection is \[O(T(R))\]

\subsection{Select-Join}
Again, since a \textit{selection} operation is performed on the mapper task only, we can join it with a \textit{join} operation easily.
We will look at the case of $\sigma_C(R(A,B)) \Join S(B,D)$.
While a mapper task that handles tuples from relation $S$ acts like the mapper in a simple join, the mapper task that loads tuples from relation $R$ acts a little differently.
First it needs to test each tuple according to condition $C$, and for each tuple that passes the test it continues acting as a regular join mapper task:
For each tuple $(a,b)$ in $R$, the mapper first checks that $C((a,b))$ is true.
If so, it emits the key-value pair $(b,(R,a))$, and for each tuple $(b,d)$ in $S$ it emits $(b,(S,d))$.
The reducer acts the same as in regular join.
It receives a key $b$ together with the list of pairs\\
$[(R,a_1),(R,a_2),...,(R,a_n),(S,d_1),(S,d_2),...,(S,d_m)]$.\\
The reducer iterates over the list and joins all the tuples from $R$ with all the tuples from $S$, and emits the key $b$ with the value of list of the joined tuples:\\
$(b,[(a_1,b,c_1),(a_1,b,c_2),...(a_n,b,c_m)])$.

The pseudo-code for the map and reduce functions for select-join operation are presented in Algorithms~\ref{alg:selection-join-map}~and \ref{alg:selection-join-reduce}~respectively. 

\begin{algorithm}
	\SetKwFunction{emit}{emit}
	\KwIn {
		$t$ a single tuples from the relation $R$.\newline
		$a,b$ attributes of $R$.\newline
		$\theta$ a binary operation.
	}
	\KwResult{
		A key-value pair, where the key does not matter, and the value is the input tuple.
	}
	\If{$\theta(t[a],t[b])$}{
		\emit($t$, $\_$)\;
	}
	\caption{The Mapper of the selection-join operation.}
	\label{alg:selection-join-map}
\end{algorithm}

\begin{algorithm}
	\SetKwFunction{Emit}{emit}

	\KwIn{Key $b$, List$<$Tagged-Value$> [(R,t_{r_1}),..,(R,t_{r_n}),(S,t_{s_1}),..,(S,t_{s_m})]$}
	\BlankLine
	\ForAll{$(\_,t_{r_i}) \in [(R,t_{r_1}),...,(R,t_{r_n})]$}{
		\ForAll{$(\_,t_{s_j}) \in [(S,t_{s_1}),...,(S,t_{s_m})]$}{
			\Emit{$b, t_{r_i} \cup t_{s_j}$}
		}
	}

	\caption{The Reducer of the selection-join operation.}
	\label{alg:selection-join-reduce}
\end{algorithm}

\subsubsection{Communication cost}
Each chunk of the relations $R$ and $S$ is fed into one map task, so the sum of the communications costs for all the map tasks is $T(R)+T(S)$
As before, the sum of outputs of the map tasks is roughly as large as their input.
So, the communication cost of a join is 
\[O(T(R)+T(S))\]

The above cost estimates will be used to compute the costs of the various plans.

%% file: probabilistic/tex/prob_example.tex
\section{Example}
\label{sec:example}
The following is an example of a distributed probabilistic database, denoted $\dbpe$.
This example will be used throughout this chapter.
The schema of the database is comprised of three relations: 
employees, departments and ranks, denoted \texttt{Emp}, \texttt{Dept}, and \texttt{Rank} respectively.
We assume that each relation resides on a different node, thus making it distributed.
Each tuple is assigned a tuple-id.
This tuple-id is used for easy reference for the tuple, but more importantly it is used as the random variable representing the probability event associated with the tuple.
Also, each tuple $t$ is associated with a probability, denoted $p(t)$.
Table~\ref{tab:prob.db.example} depicts the database.

\begin{table}[ht]
	\centering
	\captionsetup[subfloat]{margin=0cm}
	\subfloat[][]{
		\vtop{%
		  \vskip0pt
		  \hbox{%
				\begin{tabular}{ r|c|l|c|c|c }
					\multicolumn{6}{c}{\texttt{Emp}} \\
					\cline{2-5}
					   & \textbf{eid} & \textbf{name} & \textbf{did} & \textbf{rid}  \\
					\cline{2-5}
				  $e_1$ & 1 & john & 1 & 1 & 0.6 \\
				  $e_2$ & 2 & mary & 1 & 1 & 0.9 \\
				  $e_3$ & 3 & joe  & 2 & 1 & 0.8 \\
				  \cline{2-5}
				\end{tabular}
		  }
		}
	} \\
	\qquad
	\subfloat[][]{
		\vtop{%
		  \vskip0pt
		  \hbox{%
				\begin{tabular}{ r|c|l|c }
					\multicolumn{4}{c}{\texttt{Dept}} \\
					\cline{2-3}
					& \textbf{did} & \textbf{dname}  \\
					\cline{2-3}
				  $d_1$ & 1 & R\&D  & 0.5 \\
				  $d_2$ & 2 & Sales & 0.2 \\
				  \cline{2-3}
				\end{tabular}
		  }%
		}
	}
	\qquad
	\subfloat[][]{
		\vtop{%
		  \vskip0pt
		  \hbox{%
				\begin{tabular}{ r|c|l|c }
					\multicolumn{4}{c}{\texttt{Rank}} \\
					\cline{2-3}
					& \textbf{rid} & \textbf{name}  \\
					\cline{2-3}
				  $r_1$ & 1 & A & 0.5 \\
				  $r_2$ & 2 & B & 0.1 \\
				  \cline{2-3}
				\end{tabular}
			}
		}
	}
\caption{
	The three tables that comprise the example probabilistic database $\dbpe$.
	The column on the left that is out-of-the-frame contains the tuple ids, e.g. $e_1$, $e_2$, etc which represent the probabilistic event, or provenance.
	The column on the right that is out-of-the-frame contains the probabilities of the tuples, e.g. $p(e_1)=0.6$, $p(e_2)=0.9$ etc.
}
\label{tab:prob.db.example}
\end{table}

Some of the possible worlds of $\mathcal{D}_e^p$, denoted $pwd(\mathcal{D}_e^p)$ are shown in Table~\ref{tab:prob.db.possible.worlds}.
\begin{table}
\centering
\subfloat[][]{
\begin{tabular}{ |l|c| }
	\hline
	\textbf{world} & \textbf{prob}  \\
	\hline
  $\db_1=\{e_1,e_2,e_3,d_1,d_2,r_1,r_2\}$ & 0.002 \\
  $\db_2=\{e_1,e_2,e_3,d_1,d_2,r_1\}$     & 0.019 \\
  $\db_3=\{e_1,e_2,e_3,d_1,d_2,r_2\}$     & 0.002 \\
  ... &  \\
  $\db_{127}=\{r_2\}$                     & 0.000096 \\
  $\db_{128}=\phi$                        & 0.00144 \\
  \hline
\end{tabular}
}
\caption{
	Some of the $2^7$ possible worlds for $\dbpe$.
	Every possible world is denoted $\db_i$, and it contains some of the tuples, denoted by their tuple id.
	The probability associated with each of the possible worlds is a multiplication of its elements.
	For example, $\db_1$ contains all the tuples, so its probability is a simple multiplication of all the probabilities associated with all the tuples.
	$\db_2$ does not contain tuple $r_2$, so the probability is a multiplication of all the tuples except for $r_2$, multiplied by $1-p(r_2)$.
}
\label{tab:prob.db.possible.worlds}
\end{table}
\subsection{A Simple Query}
\label{subsec:a_simple_query}
Consider the following simple query, denoted $q_e$, over $\mathcal{D}_e^p$:
\begin{center}
	"\textit{Retrieve all department ids and rank ids of all the employees, such that you take into consideration the probability of the existence of the departments, ranks and employees}".
\end{center}
The formalization of the query SQL can be viewed in Figure~\ref{fig:prob.db.example.sql}.

\begin{figure}[]
\centering
\begin{minted}[baselinestretch=1,linenos=true]{sql}
SELECT
	DISTINCT
	e.did AS 'Dept ID',
	e.rid AS 'Rank ID'
FROM 
	Emp  AS e,
	Dept AS d,
	Rank AS r
WHERE 
	e.did = d.did AND
	e.rid = r.rid
\end{minted}
\caption{
	SQL formalization of the example query, $q_e$.
	Notice the \mintinline{sql}{DISTINCT} keyword: the query should retrieve only the different combination of department id and rank id.
	If two employees in the same department have the same rank than only a single row should appear in the result relation.
	Also note that although in a non-probabilistic database there was no need to do the join of the three relations, because the information is contained in the \texttt{Emp} relation.
	Here we need to do the join to make sure the probabilities are correct.
}

\label{fig:prob.db.example.sql}
\end{figure}

The query is a natural join between all three relations in $\dbpe$, and the result relation can be viewed in Table~\ref{tab:prob.db.simple.query.result}.
Note that although in a non-probabilistic database there was no need to do the join of the three relations, because the information is contained in the \texttt{Emp} relation.
Here we need to do the join to make sure the probabilities are correct.

\begin{table}
\centering
\begin{tabular}{ |c|c|c| }
	\hline
	\textbf{did} & \textbf{rid} & \textbf{p}  \\
	\hline
	1 & 1 & 0.24 \\
	2 & 1 & 0.016 \\
  \hline
\end{tabular}
\caption{
	Result relation of execution the example query, $q_e$, on the example database, $\dbpe$.
	For the first row, we sum the probabilities of all the possible worlds where either $e_1$ or $e_2$ appear, and both $d_1$ and $r_1$ appear.
}
\label{tab:prob.db.simple.query.result}
\end{table}

The probability associated with each tuple in the result relation is the sum of all the possible worlds that return that tuple when querying them with $q_e$.
Notice that this result is calculated directly from the semantics of the query.
An execution plan is not yet being considered.

\subsection{A non-Safe Execution Plan}
We now compute the result of $q_e$ over a distributed setting using the \mr~relational algebra operations not by applying it to each possible world, but by employing the following execution plan:

\begin{center}
$\mathcal{P}_1=\Pi$ \subscriptt{did,rid}
((\texttt{Rank} $\times$ \texttt{Dept}) $\bowtie$ \subscriptt{did,rid} \texttt{Emp})	
\end{center}

\subsubsection{First Step: Join}
The first step in the execution plan is a join between \texttt{Rank} and \texttt{Dept} tables.
We started with this join rather than a join that includes \texttt{Emp} due to an efficiency factor: this is a star join, and it is expected that \texttt{Emp} is very large, while \texttt{Rank} and \texttt{Dept} are significantly smaller.
We denoted the resulting join \texttt{R'}.
Notice that $T(\texttt{R'})=T(\texttt{Rank}) \cdot T(\texttt{Dept})$.
$T(\texttt{R'})$ is presented in Table~\ref{tab:prob.db.non-safe.1st.join}.

Since this step is a join, and we assume the tuples in the relations are independent probability-wise, the probabilities of the tuples in the result relation are a multiplication of the probability of the original tuples that are the source to the result tuple.

\subsubsection{Second Step: Natural Join}
The next step is a join between \texttt{R'} and \texttt{Emp}.
Since this step is again, a join, the probabilities of the tuples in the result relation are a multiplication of the probability of the original tuples that are the source to the result tuple.

The resulting relation is denoted \texttt{R''}.
Notice that $T(\texttt{R''})=T(\texttt{Emp})$.
\texttt{R''} is presented in Table~\ref{tab:prob.db.non-safe.2nd.join}.

\subsubsection{Third Step: Projection}
The final step is a projection of attributes \texttt{did} and \texttt{rid} on \texttt{R''}.
In a projection operation, tuples that have identical values in the projected attributes are merged into a single tuple.
In this case, the calculation of the probability of the result tuple is the probability of the event that at least one of the original tuples had appeared.
For example, consider two tuples $t_1$ and $t_2$.
If these two tuples are merged in a projection operation, the probability if the result tuple is:
$1-(1-p(t_1))*(1-p(t_2))$.

The result is presented in Table~\ref{tab:prob.db.non-safe.projection}. 

\subsubsection{The Plan is Wrong}
When examining the result in Table~\ref{tab:prob.db.non-safe.projection}, it can be seen that the probability of the tuple (1,1) is different from the probability calculated from by applying the query on every possible world, which means that the execution plan is faulty, or \textit{non-safe}.

% intermediate results of the non-safe plan.
\begin{table}
\centering
	\subfloat[][The result relation of \texttt{Rank} $\bowtie$ \texttt{Dept}, denoted \texttt{R'}.]{
		\begin{tabular}{ |c|l|c|c|c| }
			\hline
			\textbf{did} & \textbf{dname} & \textbf{rid} & \textbf{rname} & \textbf{probability}  \\
			\hline
			1 & R\&D  & 1 & A & 0.25 \\
			1 & R\&D  & 2 & B & 0.05  \\
			2 & Sales & 1 & A & 0.1  \\
			2 & Sales & 2 & B & 0.02  \\	
		  \hline
		\end{tabular}
		\label{tab:prob.db.non-safe.1st.join}
	}
	
	\subfloat[][The result relation of \texttt{R'} $\bowtie$ \subscriptt{did,rid} \texttt{Emp}, denoted \texttt{R''}.]{
		\begin{tabular}{ |c|l|c|l|c|c|c| }
			\hline
			\textbf{eid} & \textbf{name} & \textbf{did} & \textbf{dname} & \textbf{rid} & \textbf{rname} & \textbf{probability} \\
			\hline
			1 & john & 1 & R\&D  & 1 & A & 0.150 \\
			2 & mary & 1 & R\&D  & 1 & A & 0.225 \\
			3 & joe  & 2 & Sales & 1 & A & 0.08  \\
		  \hline
		\end{tabular}
		\label{tab:prob.db.non-safe.2nd.join}
	}

	\subfloat[][The result relation of\\$\Pi$ \subscriptt{did,rid} \texttt{R''}, which is the final\\result of the query.]{
		\begin{tabular}{ |c|c|c| }
			\hline
			\textbf{did} & \textbf{rid} & \textbf{probability}  \\
			\hline
			1 & 1 & 0.34125 \\
			2 & 1 & 0.08    \\
		  \hline
		\end{tabular}
		\label{tab:prob.db.non-safe.projection}
	}
	
	\caption{
	Step-by-step execution of the \textbf{non-safe} plan $\mathcal{P}_1$
	}
\end{table}

\subsection{A Safe Plan}
We now again execute query $q_e$, but this time we follow an execution plan that was devised using the Dalvi-Suciu algorithm~\cite{DalviSuciu:2007}:

\begin{center}
	$\mathcal{P}_2=$
	($\Pi$ \subscriptt{did,rid} (\texttt{Emp}))
	$\bowtie$ \subscriptt{did}
	($\Pi$ \subscriptt{did} (\texttt{Dept}))
	$\bowtie$ \subscriptt{rid} 
	($\Pi$ \subscriptt{rid} (\texttt{Rank}))
\end{center}

\subsection{First Step: Projections}
First, we perform three projections on the three tables.
The order of the projections is not of importance since they are independent of each other.
Without the loss of generality we first perform $\Pi$ \subscriptt{did,rid} (\texttt{Emp}).
We denote the output $\texttt{R}_1$, and $T(\texttt{R}_1)=V(\texttt{Emp},[\texttt{did},\texttt{rid}])$.
The following projection is $\Pi$ \subscriptt{did} (\texttt{Dept}).
The result relation is denoted $\texttt{R}_2$.
The last projection is $\Pi$ \subscriptt{rid} (\texttt{Rank}).
We denote the result relation $\texttt{R}_3$.

\subsubsection{Second Step: Join}
Next, we perform the join of two of the output relations for the previous step: $\texttt{R}_1$ $\bowtie$ \subscriptt{did} $\texttt{R}_2$.
We denote the output $\texttt{R}_4$, and $T(\texttt{R}_4) = V(\texttt{Emp},[\texttt{did},\texttt{rid}])$.

\subsubsection{Third Step: Join}
Finally we perform the join $\texttt{R}_4$ $\bowtie$ \subscriptt{rid} $\texttt{R}_3$.

As can be seen from Table~\ref{tab:prob.db.safe}, this plan is safe and its values match the values obtained using the possible world definition in Table~\ref{tab:prob.db.example}.

% intermediate results of the safe plan.
\begin{table}[ht]
	\centering
	\captionsetup[subfloat]{margin=0cm}
	\subfloat[][Relation $\texttt{R}_1$, the result of $\Pi$ \subscriptt{did,rid} (\texttt{Emp})]{
		\vtop{%
			\vskip0pt
			\hbox{%
			\begin{tabular}{ |c|c|c| }
				\hline
				\textbf{did} & \textbf{rid} & \textbf{probability}  \\
				\hline
				1 & 1 & 0.96 \\
				2 & 1 & 0.8  \\
				\hline
			\end{tabular}
			\label{tab:prob.db.safe.R1}
			}
		}
	}

	\qquad
	\subfloat[][Relation $\texttt{R}_2$, the result of $\Pi$ \subscriptt{did} (\texttt{Dept})]{
		\vtop{%
			\vskip0pt
			\hbox{%
			\begin{tabular}{ |c|c| }
				\hline
				\textbf{did} & \textbf{probability}  \\
				\hline
				1 & 0.5  \\
				2 & 0.2  \\
				\hline
			\end{tabular}
			\label{tab:prob.db.safe.R2}
			}%
		}
	}
	\qquad
	\subfloat[][Relation $\texttt{R}_3$, the result of $\Pi$ \subscriptt{rid} (\texttt{Rank})]{
		\vtop{%
			\vskip0pt
			\hbox{%
			\begin{tabular}{ |c|c| }
				\hline
				\textbf{rid} & \textbf{probability}  \\
				\hline
				1 & 0.5  \\
				2 & 0.1  \\
				\hline
			\end{tabular}
			\label{tab:prob.db.safe.R3}
			}
		}
	}

	\subfloat[][Relation $\texttt{R}_4$, $\texttt{R}_1$ $\bowtie$ \subscriptt{did} $\texttt{R}_2$]{
	\begin{tabular}{ |c|c|c| }
		\hline
		\textbf{did} & \textbf{rid} & \textbf{probability}  \\
		\hline
		1 & 1 & 0.48  \\
		2 & 1 & 0.16  \\
		\hline
	\end{tabular}
	\label{tab:prob.db.safe.R4}
}

\subfloat[][Relation $\texttt{R}_5$, $\texttt{R}_4$ $\bowtie$ \subscriptt{rid} $\texttt{R}_3$]{
	\begin{tabular}{ |c|c|c| }
		\hline
		\textbf{did} & \textbf{rid} & \textbf{probability}  \\
		\hline
		1 & 1 & 0.24 \\
		2 & 1 & 0.08 \\
		\hline
	\end{tabular}
	\label{tab:prob.db.safe.R5}
}

\caption{Step-by-step execution of the \textbf{safe} plan $\mathcal{P}_1$. \label{tab:prob.db.safe}}

\end{table}

%% file: probabilistic/tex/prob_opt.of.safe.and.non-safe.tex
\section{Communication Cost and Optimization of Safe and Non-Safe Execution Plans}
\subsection{Communication Cost of Plans}
First, we review the communication cost of the plans we have just described.

\subsubsection{Non-safe Plan}
The plan has three steps: a \textit{join} step, a \textit{natural join} step and a \textit{projection} step.
The communication cost of the first step is $T(\texttt{Rank})+T(\texttt{Dept})$, which is the number of input tuples.
The communication cost of the second step is
$T(\texttt{Rank}) \cdot T(\texttt{Dept}) + T(\texttt{Emp})$.
The communication cost of the last step is $T(\texttt{Emp})$.

The total communication cost of the execution plan is:
\begin{center}
	$cost(\mathcal{P}_1)=$
	$T(\texttt{Rank}) + T(\texttt{Dept}) + T(\texttt{Rank}) \cdot T(\texttt{Dept}) + 2T(\texttt{Emp})$
\end{center}

\subsubsection{Safe Plan}
The safe plan has also three steps.

The first step is composed of three independent projection operations.
Their communication cost is $T(\texttt{Emp})$, $T(\texttt{Dept})$ and $T(\texttt{Rank})$ respectively.

The communication cost of the second step, which is a join, is $T(\texttt{R}_1) + T(\texttt{R}_2)$, which is $V(\texttt{Emp},[\texttt{did},\texttt{rid}]) + T(\texttt{Dept})$.

The communication cost of the final step which is also a join is 
$V(\texttt{Emp},[\texttt{did},\texttt{rid}]) + T(\texttt{Rank})$.

The total communication cost is
\begin{center}
	$cost(\mathcal{P}_2)=$
	$T(\texttt{Emp}) + 2 \cdot T(\texttt{Dept}) + 2 \cdot T(\texttt{Rank}) + 2 \cdot V(\texttt{Emp},[\texttt{did},\texttt{rid}])$
\end{center}

Note that by following the Afrati-Ullman optimization~\cite{AfratiUllman2010} and performing a 3-way join instead 2 binary-joins we can get even better performance.

\subsection{Comparing the Communication Costs}
\label{subsec:compating-cc}
We now explore the conditions when $cost(\mathcal{P}_2) < cost(\mathcal{P}_1)$.
This condition is true in a database instance where
$T(\texttt{Emp}) > T(\texttt{Dept}) + T(\texttt{Rank}) + V(\texttt{Emp},[\texttt{did},\texttt{rid}])$.
This condition holds on most real-world databases: the employee table contains many more records than the departments and ranks tables, but not on all of them.

This means that the cost of the plans depends on the instance!
On some instances $\mathcal{P}_1$ is faster, and on other instance $\mathcal{P}_2$ is faster.

%% file: probabilistic/tex/prob_safe.plan.opt.tex
\section{Safe Plan Optimization}

\subsection{Motivation}
\begin{theorem}
	\label{thm:safe.plan.is.optimal}
%	For every star-join-safe-query $q$ there exists a database instance $i$ and a \textbf{safe} binary-join plan $\mathcal{P}$ such that $cost(i,\mathcal{P})$ is optimal.
	For every star-join safe query $q$ there exists a safety-preserving execution plan $\mathcal{P}_q$ such that the cost of $\mathcal{P}_q$ is optimal.
\end{theorem}

The proof for this theorem is constructive, and is given in Section \ref{subsec:plan-optimization-algo}.

\begin{theorem}
	\label{thm:non.safe.plan.is.optimal}
%	For every star-join-safe-query $q$ there exists a database instance $i$ and a \textbf{non-safe} binary-join plan $\mathcal{P}$ such that $cost(i,\mathcal{P})$ is optimal.
	There exists star-join safe query $q$ and a database instance $i$ for which there exists a non-safe binary plan $\mathcal{P}_q$ such that the cost of $\mathcal{P}_q$ is optimal.
\end{theorem}

The proof of this theorem was shown by the example in Section \ref{subsec:compating-cc}.

These theorems mean that among all the safety preserving plans, there is one which is optimal, and our algorithm can find it, but there may be plans which do not preserve safety that may have a smaller communication cost.

Theorems~\ref{thm:safe.plan.is.optimal} and~\ref{thm:non.safe.plan.is.optimal} also prove that using traditional optimizers for optimizing a \mr~plan can be harmful: depending on the database instance, the result plan is sometimes safe and sometimes not safe.
Therefore a new optimizer, which guarantees safety, is needed.

\subsection{The MRQL Optimization is not Always Safe}
Some work has been done in the area of \mr~query optimization.
The most promising work is of Fegaras et al from 2012~\cite{fegaras2012optimization}, which is now an Apache Incubator project ~\cite{mrql:apache}.
In their paper they present an optimization framework for \mr~queries.
Their work is actually more extensive: they present a new query language with features like sub-select in the group-by and select sections, but we are interested in their optimizer.
The input is an MRQL query (similar to SQL), and the output is a \mr~based execution plan.
The output plan is optimized in terms of \mr, however some issues arise:
\begin{enumerate}
	\item Their measure of optimization is the number \mr~jobs and not the communication-cost.
	
	\item 
		The input is not an SQL statement but rather an MRQL statement.
		MRQL is very similar to SQL, but still a minor conversion is required.
		
	\item Improves mainly queries with group-by statements or sub-queries.
	
	\item The execution plan is not always safe.
\end{enumerate}
We have translated the example query into the following MRQL query:
\\
\begin{center}
	\begin{minted}[baselinestretch=1,linenos=true]{sql}
	SELECT
		distinct d.did, r.rid
	FROM
		e IN emp,
		d IN dept,
		r IN rank
	WHERE e.did=d.did 
	  AND e.rid=r.rid;
	\end{minted}	
\end{center}

The optimization framework produces the following optimized execution plan:
\begin{verbatim}
MapReduce2:
    left: MapReduce2:
        left: Source (line): "queries/emp.txt"
        right: Source (line): "queries/dept.txt"
    right: Source (line): "queries/rank.txt"
\end{verbatim}
The result plan is composed of two \mr~jobs.
The first \mr~job is a join between \texttt{Emp} and \texttt{Dept} tables, and the following job is a join between the result of the first one with \texttt{Rank} table.
As we have shown earlier, the join between \texttt{Emp} and \texttt{Dept} tables makes the plan a non-safe one.
We have used the framework on several more queries, all produced a non-safe plan.

The fact that MRQL does not necessarily generate a safe-plan, is another major component in the motivation for this work that will provide a framework, or algorithm, for generating an always safe, and usually optimal plans.

\subsection{Safe-Preserving Algebraic Laws}
When we imply the Dalvi-Sucio algorithm we obtain one possible safe plan.
The next step is to rewrite the plan using safe-preserving algebraic laws.
However, safe-preserving algebraic laws are different from traditional algebraic laws.
For instance, $\Pi$ and $\bowtie$ do not commute.
Dalvi and Sucio list the following safe-preserving transformation rules:
\begin{enumerate}
	\item 
		\textbf{Pushing Selections Bellow a Join}\\
		If all the attributes in condition $c$ involve attributes only from one expression being joined (say $R$) then		
		\[\sigma_c (R \bowtie S) \Leftrightarrow \sigma_c(R) \bowtie S\]
		If condition $c$ is a conjunction of two conditions, $c_1$ and $c_2$ such that $c_1$ involves only the attributes of $R$ and $c_2$ involves only the attributes of $S$ then
		\[\sigma_{c_1 \wedge c_2} (R \bowtie S) \Leftrightarrow \sigma_{c_1}(R) \bowtie \sigma_{c_2}(S)\]
	\item 
		\textbf{Pushing Selections Bellow a Projection}
		\[\sigma_c(\Pi_a(R)) \Leftrightarrow \Pi_a(\sigma_c(R))\]
	\item 
		\textbf{Join Commutativity} Joins are commutative:
		\[R \bowtie S \Leftrightarrow S \bowtie R\]
	\item
		\textbf{Join Associativity} Joins are associative:
		\[R \bowtie (S \bowtie T) \Leftrightarrow (R \bowtie S) \bowtie T\]
	\item
		\textbf{Cascading Projections} 
		Successively eliminating attributes from a relation is equivalent to simply eliminating all but the attributes retained by the last projection:
		\[\Pi_A(\Pi_{A \cup B}(R)) \Leftrightarrow \Pi_A(R)\]
	\item
		\textbf{Pushing Projections Below a Join} A projection can be pushed below a join if it retains all the attributes used in the join:
		\[\Pi_A(R \bowtie S) \Rightarrow (\Pi_{A_R}(R))  \bowtie (\Pi_{A_S}(S))\]
		Where $A_R = A \cap Attr(R)$ and $A_S = A \cap Attr(S)$.
		
	\item
		\textbf{Lifting Projections Up a Join}\\
		A projection cannot always be lifted up a join.
		The following transformation rule can be applied only when the top $\Pi$ operator holds the following:
		For a query plan $p$, $\Pi_{a_1,...a_k}$ is safe in $\Pi_{a_1,...a_k}(p)$ if and only if for every $R \ in Rels(p)$ the following can be inferred from $\Gamma(p)$:\\
		$a_1,...,a_k,R.E \rightarrow Head(p)$
		$(\Pi_A(R)) \bowtie S \Rightarrow \Pi_{A \cup Attrs(S)}(R \bowtie S)$
\end{enumerate}

\subsection{Estimating Cost of a Plan}
In order to choose the most efficient plan, we need to know the cost of the various plans.
We cannot know these costs exactly without executing the plan.
However, most times, the cost of executing a plan is significantly greater than all the work done by the query optimizer in selecting a plan.
Therefore, we do not want to execute more than one plan for one query, and we are forced to estimate the cost of any plan without executing it.

For the estimations we require some information regarding the relations in the database, and we use well known notations from~\cite{ullman2002database}:
\begin{itemize}
	\item $T(R)$ the number of tuples in relation $R$.
	\item $V(R,\alpha)$ the value-count for attribute $\alpha$ of relation $R$, i.e. the number of distinct values relation $R$ has in column $\alpha$.
\end{itemize}

So, our problem is how to estimate costs of plans accurately.
Given values for these parameters, we may make a number of reasonable estimates of relations sizes that can be used to predict the cost of a complete execution plan.

\subsubsection{Estimating Size of a Intermediate Relations~\cite{Korth:1986:DSC:6204}}
For each operator we will give an upper-bound and an estimation on the size of the result relation.

\textbf{Projection}
The projection operation eliminates duplicate values.
Let $S=\Pi_{a_1,a_2,...a_n}(R)$, then $T(S)=V(R,[a_1,a_2,...a_n])$.
However, often we do not have this statistic available, so it must be approximated.
In the extreme, $T(S)=T(R)$, or as small as $1$.
Another upper limit on $T(S)$ is the maximum number of distinct tuples that could exist:
$\prod_{i=1..n}V(R,a_i)$.

The upper bound is:
\[T(S)=min(T(R),\prod_{i=1..n}V(R,a_i))\]

The acceptable estimation is:
\[T(S)=min(T(R)/2,\prod_{i=1..n}V(R,a_i))\]

\textbf{Selection}
When we perform a selection, we almost always reduce the number of tuples.
The upper bound is the case all tuples hold the constraint and therefore:
\[T(S)=T(R)\]
Let $S=\sigma_{A=c}(R)$, where $A$ is an attribute of $R$, and $c$ is a constant.
In the simplest kind of selection, where an attribute is compared to a constant, there is an easy way to estimate the result, provided we know, or can estimate, the number of different values the attribute has.
Then we can estimate:
\[T(S)=T(R)/V(R,A)\]
The size estimate is more problematic when the selection involves an inequality comparison, for instance $S=\sigma_{A<c}(R)$.
If we know the value of $c$ and the lowest and highest values of attribute $A$ during the time of cost estimation, we can make the following estimation(cite Silberschatz):
\[T(S)=\frac{c-min(A,R)}{max(A,R)-min(A,R)}\]
However, we shall assume the worst case when we don't know the value of $c$ (e.g. $c$ is a parameter).
Ullman[cite] proposes that the typical inequality will return about one third of the tuples (Silberschatz says half of the tuples):
\[T(S)=T(S)/3\]
In the case of a "not equal" comparison, we can assume that most tuples will satisfy the condition.
That is, for the selection $S=\sigma_{A \neq c}(R)$ we can estimate:
\[T(S)=T(R)\]

% \todo{handle AND and OR}

\textbf{Join}
We shall consider here only equi-joins (also include natural-joins).
Theta-joins can be estimated as if they were a selection following a product.
For start, we shell assume that the equi-join of two relations involves only the equality of two attributes, one of each relation.
i.e., we consider the join $R(A,B) \times_{B=C} S(C,D)$.
The upper bound is simply a size of a Cartesian product:
\[T(R \times S)=T(R) \cdot T(S)\]
We shell look at this estimation problem as a probability problem:
Suppose $r$ and $s$ are tuples in $R$ and $S$ respectively.
What is the probability of $r$ and $s$ agree on the joint attributes?
Suppose that $V(R,B) \geq V(S,C)$.
Then $C$-value is surely one of the $B$-value that appear in $R$.
Hence, the chance that $r$ is joined with $s$ is $1/V(R,B)$.
Similarly, if $V(R,B) < V(S,C)$, then the chance that $r$ and $s$ will match is $1/V(S,C)$.
In general, the probability of $r$ matching $s$ is $1/max(V(R,B),V(S,C))$. Thus:
\[T(R \times S)=T(R) \cdot T(S)/max(V(R,B),V(S,C))\]

\subsection{Plan Cost Estimation Algorithm}
An execution plan is a relational algebra expression.
As such, it can be considered a tree which its vertices are relational algebra operators, and its leaves are relations.
We make a distinction between a \textit{logical plan} and a \textit{physical plan}.
A logical plan is a relational algebra expression that describes the relational operators needed to be taken to execute the query.
A physical plan is a translation of the logical plan into the \mr~world.
The physical plan describes exactly which \mr~actions to take in order to execute the plan.
The importance of the physical plan comes from section~\ref{sec:rel.alg.using.map.reduce}: 
a single relational operator is usually mapped into a single \mr~job, however in order to lower the communication cost we can combine two relational operators and map them to a single \mr~job.
We aspire to find the most efficient physical plan to a given logical plan in terms of communication cost.

Algorithm~\ref{alg:opt.physical.plan} does so.
This algorithm is based on the query optimization Algorithm 13.7 in \cite{silberschatz:Database} (page 600), except that the physical plan is a \mr~job.
The input is a logical plan, in the form of a tree.
The output is the optimal physical plan in terms of communication cost together with its result size and estimated communication cost.
Since this function will be called repeatedly, we can develop a dynamic programming algorithm for estimating the cost and size of plans.
Dynamic programing algorithms store results of computations and reuse them, a method that can greatly reduce the execution time.
The function stores the cost and size of plans in an associative array $mem$, which is indexed by logical plans.

% cost algorithm
\begin{function}
	\caption{optPhyPlan()\label{alg:opt.physical.plan}}
	\LinesNumbered
	% functions
	\SetKwFunction{isLeaf}{isLeaf}
	\SetKwFunction{Match}{match}
	\SetKwFunction{Subplan}{subplan}
	\SetKwFunction{Cost}{cost}
	\SetKwFunction{Min}{min}

	$PATTERNS$ is a set of relational expressions that can be mapped into a single map-reduce job.\;
	$subplan(p, \mathcal{P})$ is a function that returns the sub plan of $\mathcal{P}$ after matching it with pattern $p$.\;
	$cost(p,s)$ is a function that return the cost of a pattern $p$ given the input size $s$.\;
	$mem$ is an array that contains results of $subplan$.\;
	
	\KwIn{A logical plan $\mathcal{P}$}
	\KwOut{The optimal physical plan with its estimated communication cost and result size}
	\BlankLine
	\If{mem[$\mathcal{P}$] $\neq -1$}{
		\KwRet{(mem[$\mathcal{P}$])}
	}	
	\If{\isLeaf{$\mathcal{P}$}}{
		\{ a triplet consists of plan, cost and size\}\;
		\KwRet{($\phi$,0,$\mathcal{P}$.size)} 
	}
	possiblePlans $\leftarrow$ $\phi$\;
	\For{$p \in PATTERNS$} {
		\If{\Match($p$, $\mathcal{P}$)} {
			optsubPlan $\leftarrow$ \optPhyPlan(\Subplan(p,$\mathcal{P}$))\;
			possiblePlans.add(optsubPlan, optsubPlan.cost + \Cost(p,optsubPlan.size)\;
		}
	}
	mem[$\mathcal{P}$] $\leftarrow$ \Min(possiblePlans)\;
	\KwRet{mem[$\mathcal{P}$]}	
\end{function}

\subsection{Improving a Safe Plan}
From the classic database literature, there are a number of algebraic laws that tend to improve logical query plans, and are commonly used in optimizers.
We will review them first, and then make the appropriate modifications for safe-preservation:
\begin{itemize}
	\item 
		Selections can be pushed down the expression tree as far as they can go.
		If a selection condition is the AND of several conditions, then we can split the condition and push each piece down the tree separately.
	\item
		Projections can be pushed down the tree, or new projections can be added.
		We may introduce a projection as long as it eliminated only attributes that are neither used by an operator above nor are in the result of the entire expression.
	\item
		Certain selections can be combined with a product below to turn the pair of operations into an equi-join, which is generally much more efficient to evaluate.
\end{itemize}

\subsection{Plan Optimization Algorithm}
\label{subsec:plan-optimization-algo}
Given a query $q$, we first need to generate a safe execution plan for it.
We generate the plan using Algorithm~\ref{alg:gen.safe.plan} which is based on the paper by Dalvi and Suciu~\cite{DalviSuciu:2007}, and denote the safe plan $\mathcal{P}$.
Given a safe plan $\mathcal{P}$, we need to generate other safe plans that are equivalent to it.
An exhaustive search algorithm would be to traverse all sub plans of a plan, then, using the equivalence rules, replace that sub plan with an equivalent one.
This process goes as follows.
Given an execution plan $\mathcal{P}$, the set of equivalent plans, $\mathcal{EP}$, initially contains only $\mathcal{P}$.
Now, each plan in $\mathcal{EP}$ is matched with each equivalence rule.
If a plan, say $\mathcal{P}'$, matches one side on an equivalence rule then the optimizer applies the rule and generates a new plan.
The new plan is added to $\mathcal{EP}$.
This process continues until no more new plans can be generated.
Algorithm~\ref{alg:gen.all.equiv.plans} outlines this process.

\begin{function}	
	\caption{safePlan($q$)\label{alg:gen.safe.plan}}
	
	\RestyleAlgo{algoruled}
	
	\KwIn{$q$ query}
	\KwResult{$\mathcal{P}$ safe plan if exists}
	\BlankLine
	
	\If{$Head(q)=Attr(q)$}{
		\Return{Any plan $\mathcal{P}$ for $q$}
	}

	\For{$A \in (Attr(q) - Head(q))$}{
		let $q_A$ be the query obtained from $q$ by adding $A$ to the head variables.
		\If{$\Pi_{Head(q)}(q_A)$ is a safe operator}{
			\Return{$\Pi_{Head(q)}(safePlan(q_A))$}
		}
	}
	
	Split $q$ into $q_1 \bowtie_c q_2$, s.t.
	$\forall R_1 \in Rels(q_1), R_2 \in Rels(q_2)$, $R_1$ and $R_2$ are separated
	
	\If{No such split exists}{
		\Return{"No safe plan exists"}
	}

	\Return{$safePlan(q_1) \bowtie_c safePlan(q_2)$}

\end{function}

\begin{procedure}	
	\caption{generateAllEquivalent($\mathcal{P}$)\label{alg:gen.all.equiv.plans}}

	\RestyleAlgo{algoruled}
	
	\KwIn{$\mathcal{P}$ Execution plan}
	\BlankLine
	
	$\mathcal{EP}$ $\leftarrow$ $\mathcal{P}$ \;
	\Repeat{no new plans can be added to $\mathcal{EP}$} {
		Match each plan $\mathcal{P}_i$ in $\mathcal{EP}$ with each equivalence rule $r_j$
		\If{$r_j$ can be applied to $\mathcal{P}_i$}{
			Create a new plan $\mathcal{P}'$ by applying $r_j$
			Add $\mathcal{P}'$ to $\mathcal{EP}$ if it is not already in $\mathcal{EP}$
		}
	}	
\end{procedure}

This algorithm is extremely costly in both space and time, but it can be greatly reduced in both measures by using three key ideas:
\begin{enumerate}
	\item If we generate a plan $\mathcal{P}'$ from a plan $\mathcal{P}_1$ by using an equivalence rule on sub plan $p_i$, then $\mathcal{P}'$ and $\mathcal{P}_1$ have identical sub plans except for $e_i$ and its transformation.
	Plan-representation technique that allows both plans to point to shared sub plans can significatly reduce the needed space.
	\item It is not always necessary to generate every expression that can be generated with equivalence rules.
	If the optimizer takes cost estimates of evaluation into account, it may be able to avoid examining some of the plans.  
	\item Some of the rules can be proved to reduce the communication cost of a plan.
	This means that we can apply these rules without further checking, and the new plan is not worse than the original plan.
	We elaborate on this idea below.
\end{enumerate}

\subsection{Sure Rules}
We now examine some rules and prove that applying these rules will surely reduce the communication cost of a plan.

\subsubsection{Pushing Selection Before (below) a Join}
Consider the following plan $\mathcal{P}_1: \sigma_c(R \bowtie S)$.
The communication cost of $\mathcal{P}_1$ is 
$2 \cdot T(R) + 2 \cdot T(S) + \frac{T(R) \cdot T(S)}{max(V(R,A),V(S,A))}$.
Suppose we use equivalence rule 1 on $\mathcal{P}_1$.
We get $\mathcal{P}_2: \sigma_c(R) \bowtie S$.
In the case of $\mathcal{P}_2$, we have a join operation right after a selection operation, therefore we can use the select-join operator.
So, the communication cost of $\mathcal{P}_2$ is $T(R) + 2 \cdot T(S) + \frac{T(R)}{V(R,A)}$
As we can see, the cost of $\mathcal{P}_2$ is lesser than the cost of $\mathcal{P}_1$ on any database instance!
This means that we can use this rule without calculating the cost.
This rule will always yield an optimized plan.

\subsubsection{Pushing Selection Before (below) a Projection}
Consider the following plan $\mathcal{P}_1: \sigma_c(\pi_{\alpha}(R))$.
The communication cost of $\mathcal{P}_1$ is 
$2T(R) + T(R') + \frac{T(R')}{V(R',c)}$
where $R'$ is the result relation of $\pi_{\alpha}(R)$.
Suppose we use equivalence rule 2 on $\mathcal{P}_1$.
We get $\mathcal{P}_2: \pi_{\alpha}(\sigma_c(R))$.
In the case of $\mathcal{P}_2$, we have a projection operation right after a selection operation, therefore we can use the select-project operator.
So, the communication cost of $\mathcal{P}_2$ is $T(R) + \frac{T(R)}{V(R,c)}$
Even in the rare case of $R=\sigma_c(R)$, $\mathcal{P}_2$ is preferred over $\mathcal{P}_1$ on any instance.
Therefore we should always apply this rule, without checking the cost of the new plan.

\subsubsection{Unifying Projections}
Consider the following plan: $\mathcal{P}_1: \pi_{\alpha}(\pi_{\alpha \cup \beta}(R))$.
It is obvious that unifying the relations, i.e. projecting only $\alpha$, will yield an equivalent and more efficient plan.

\subsection{Rules that need to be Checked}
Consider the following plan: $\mathcal{P}_1: \pi_{\alpha}(R \bowtie S)$.
The communication cost of $\mathcal{P}_1$ is $2T(R) + 2T(S) + 2 \cdot \frac{T(R) \cdot T(S)}{max(V(R,A),V(R,A))}$.
Now, consider applying rule 6 on $\mathcal{P}_1$.
The result plan is $\mathcal{P}_2: \pi_{\alpha_R}(R) \bowtie \pi_{\alpha_S}(S)$.
The communication cost of $\mathcal{P}_2$ is 
$2T(R) + 2T(S) + \frac{2T(R)}{V(R,\alpha_R)} + \frac{2T(S)}{V(S,\alpha_S)}$.

Which plan is more optimized?

When we compare the communication costs of $\mathcal{P}_1$ and $\mathcal{P}_2$, we get that $\mathcal{P}_1$ will cost less if approximately $T(R) \cdot T(S) < T(R) + T(S)$, meaning that it is instance-dependent.
We cannot know which plan is better without calculating the cost of the plans.

\subsection{Finding the Optimized Plan}
An algorithm for finding the most efficient plan is as follows:
The input is a query $q$.
First, we generate a safe execution plan $\mathcal{P}$.
The algorithm is recursive, so at first the stop condition is checked, and is whether $\mathcal{P}$ is only a leaf, meaning that it has no descendant.
If this is the case than nothing can be optimized and $\mathcal{P}$ returns as is.
Otherwise, all the sure rules are applied.
Now we are left with the optional rules.
In order to know whether to apply them or not, we need check the cost with them and without them, and then make the decision.
Finally, we continue the recursion by calling it on the sub-plan.

The algorithm is presented in Algorithm \ref{alg:find.best.plan}.

\begin{function}	
	\caption{findBestPlan($q$)\label{alg:find.best.plan}}

	\RestyleAlgo{algoruled}
	
	\KwIn{$q$ query}
	\BlankLine
	
	$\mathcal{P} \leftarrow safePlan(q)$ \;
	\If{isNull?($\mathcal{P}$)}{
		\Return{"No safe plan exists"}
	}	
	
	\If{isLeaf?($\mathcal{P}$)}{
		\Return{$\mathcal{P}$}
	}
	
	applySureRules($\mathcal{P}$) \;
	
	\Switch{}{
		\uCase{check rule can be applied}{
			$\mathcal{P}' \leftarrow applyCheckRule(\mathcal{P})$ \;
			\eIf{cost($\mathcal{P}'$) $<$ cost($\mathcal{P}$)} {
				\findBestPlan{$\mathcal{P}'.subPlan$} \;
			}{
				\findBestPlan{$\mathcal{P}.subPlan$} \;
			}
		}
		\Other{
			\findBestPlan{$\mathcal{P}.subPlan$} \;
		}
	}
	
\end{function}

%% file: probabilistic/tex/prob_experiments.tex
\section{Experiments}
\subsection{Data}
\subsubsection{Real Data}
NELL~\cite{nell}, or Never-Ending Language Learning is a research project that attempts to create a computer system that learns over time to read the web.
It attempts to "read," or extract facts from text found in hundreds of millions of web pages and associates each read fact with a confidence level.
In our experiment we have extracted tables from this dataset, using the confidence as the probability for each record (or tuple).
We have extracted the following tables and used them as the input for all of the experiments:
\begin{itemize}
	\item movies \{mid,name\}, 6,844 records
	\item actors \{aid,name\}, 32,806 records
	\item movie\_actors \{mid,aid\}, 40,539 records
\end{itemize}

\subsubsection{Synthetic Data}
Since the real data is not big in the sense that a single computer cannot handle (which makes the use of the \mr~framework cumbersome) we have used the same relations to generate synthetic records. 
\begin{itemize}
	\item movies, 1m records
	\item actors, 1m records
	\item movie\_actors, 5m records
\end{itemize}

\subsection{Setup}
We run all the experiments on the Amazon Elastic \mr~\cite{amazon.emr}~infrastructure.
The real data run was executed on a single machine, and the synthetic data run was executed on four machines, each is an SSD-based instance storage for fast I/O performance with a quad core CPU and 8 GiB of memory.
All machines run \hd~version 2.6.0 with Java 8.

\subsection{Measurement}
We used communication-cost (see Section~\ref{sec:communication-cost}) as the main measurement for comparing the performance of the different algorithms.
The input records to each map task and reduce task were simply counted and summed up and the end of the execution.
This count is performed on each machine in a distributive manner.
The implementation of \hd~provides an internal input records counter that makes the counting and summing task extremely easy.
Communication-cost is an infrastructure-free measurement, meaning that it is not affected by weaker/stronger hardware or temporary network overloads, making it our measurement of choice.
However we also measured the time of execution: we run each experiment 3 times and give the average time.

\subsection{The Query}
The query we executed throughout the experiment is as follows:
\begin{center}
	"\textit{Generate a list of all the movie ids and actor ids, such that we take into consideration the probability of the existence of the movie and the actor}".
\end{center}

The following SQL query is the formalization of the requested query.
It is safe.

\begin{center}
	\begin{minted}[baselinestretch=1,linenos=true]{sql}
		SELECT
		    movie.mid,actor.aid
		FROM 
		    movies       as m, 
		    actors       as a, 
		    movie_actors as m_a
		WHERE 
		    movie_actors.mid = movies.mid AND
		    movie_actors.aid = actors.mid
	\end{minted}
\end{center}

%The plan from Suciu's paper after trivial optimizations is:
%\begin{align*}
%& \Pi_{m\_a.aid, m\_a.mid}(\\
%& ~~\Pi_{a.aname, m\_a.aid, m\_a.mid}(\\
%& ~~~~\Pi_{a.aname, m.mname, m\_a.aid, m\_a.mid}(\\
%& ~~~~~~\Pi_{a.aname, m.mname, m\_a.aid, m\_a.mid, a.aid}(\\
%& ~~~~~~~~\Pi_{mname, mid, aid, aname}(m \bowtie m\_a)))))
%\end{align*}

\subsection{Control Group}
As a control group, we use the non-safe execution plan created by MRQL, and attaching to each result tuple its provenance.
Then, we calculate the probability of each of the result tuples from the provenance.
Consider Table~\ref{tab:prob.db.provenance} for the provenance of the example query.

\begin{table}
	\centering
	\begin{tabular}{ |c|c|c }
		\cline{1-2}
		\textbf{did} & \textbf{rid} & \textbf{provenance}  \\
		\cline{1-2}
		1 & 1 & $(d_1 \wedge r_1 \wedge e_1)\vee(d_1 \wedge r_1 \wedge e_2)$ \\
		2 & 1 & $d_1 \wedge r_1 \wedge e_3$ \\
		\cline{1-2}
	\end{tabular}
	\caption{The result table of the example query, with the provenance for each tuple.\label{tab:prob.db.provenance}}	
\end{table}

\subsection{The Experiments}
For the query above, we created three execution plans:
\begin{itemize}
	\item Sucio's plan naively transformed to \mr.
	\item Our efficient \mr~plan.
	\item A non-safe plan using MRQL with provenance.
\end{itemize}
We have executed the three plans on the NELL data and the synthetic data, and measured the execution time.

\subsection{Results}

\begin{figure}
	\centering
	\includegraphics [width=10.5cm] {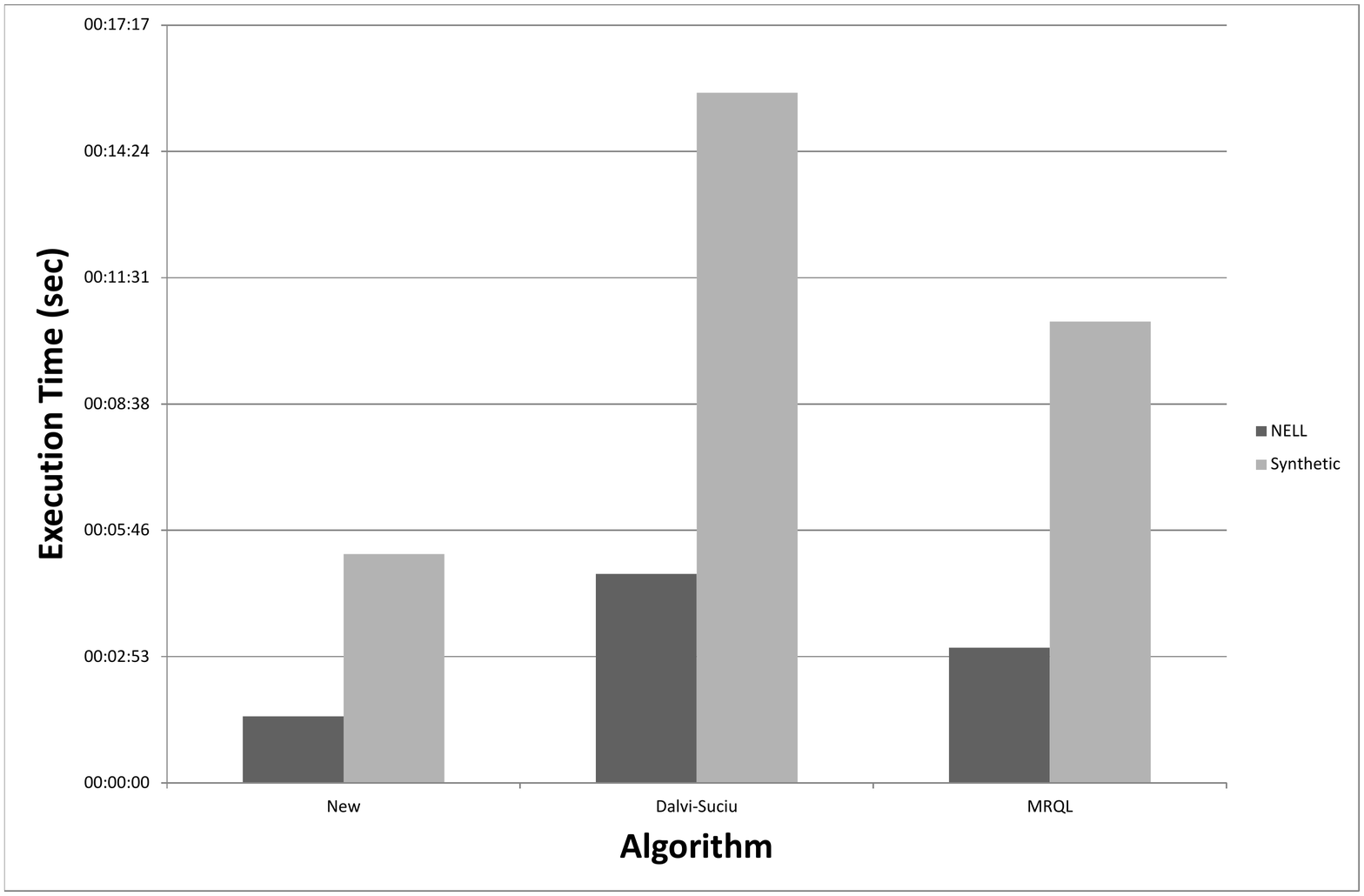}
	\caption{
		Comparing the execution time of the three plans on the synthetic and NELL (real) data.
	}
	\label{fig:prob.results}
\end{figure}

The execution times of the algorithms on the data are presented in Figure \ref{fig:prob.results}.
As can be seen, the  new algorithm outperforms the two others; however it is interesting to notice that the MRQL, which is the 'control group', is faster than the Dalvi-Suciu algorithm.
We believe that it happened due to the following reasons:
\begin{itemize}
	\item Since MRQL does not take into account plan safety, its plans are greatly more efficient.
	\item The calculation of the probability is calculated concurrently, which greatly reduces the execution time.
	\item The provenance resulting from the query is not complicated, and can be calculated easily.
\end{itemize}

%% file: probabilistic/tex/prob_conclusion.tex
\section{Conclusion}
In this chapter we have investigated the problem of querying distributed probabilistic databases.

First, we have reviewed the classic problem of querying centralized probabilistic database, and showed the Dalvi-Suciu dichotomy, which differentiates  queries that can be executed in a linear time (safe queries) from ones that cannot (non-safe queries). 

Next, we have moved to the distributed database field, and showed how to translate relational algebra operators into \mr~jobs, and we have introduced some combined operators, which will play part in optimizing queries.

Than we have presented two theorems that are the main motivation for this chapter: current optimizers for \mr~jobs may break the safety of an execution plan, but among all the safe plans we can find the most efficient one.

We have presented a search algorithm for finding the most efficient execution plan, among all the safe plans, to be executed with a series of \mr~jobs. 

We have tested this algorithm against other methods and have shown that it outperforms them.

%% file: hadoop.cache/tex/cache_intro.tex
\chapter{Universal \hd~Cache}
\label{ch:hadoop-cache}

\bookquote{
	If we have data, let's look at data. If all we have are opinions, let's go with mine.
}{Jim Barksdale, CEO of Netscape}

\section{The \hd~Distributed File System}
As mentioned earlier, \hd~comes out-of-the-box with a distributed file system called HDFS, which stands for \hd~Distributed Filesystem.
HDFS is designed for storing very large files focusing on data access by streaming.
The idea behind the design is that the most efficient data processing pattern is a write-once, read-many-times pattern.
A dataset is typically generated or imported from a source, then various analyses and processing are performed on that dataset over time.
Each analysis will involve a large proportion, if not all, of the dataset, so the time to read the whole dataset is more important than the latency in reading the first record.

In this chapter we show how \mr~ algorithms can be improved with a caching mechanism.
The mechanism is universal, meaning that it does not depend on where the data is loaded from, however some important concepts come from HDFS, therefore we start by describing it.

\subsection{HDFS Concepts}
\subsubsection{Blocks}
A block, in the context of a file system, is the minimum data unit that can be read or write.
It is usually a few kilobytes in size.
The concept of block is usually transparent to a file system user, who is simply reading or writing a file of whatever length.

HDFS, too, has the concept of a block, but it is a much larger unit-128 MB by default.

Like in a file system for a single disk, files in HDFS are broken into block-sized chunks, which are stored as independent units.

\subsubsection{Namenodes and Datanodes}
An HDFS cluster has two types of nodes, working in a master-worker paradigm: a \textit{namenode} (the master) and a number of \textit{datanodes} (workers).
The namenode manages the file system namespace.
It maintains the file system tree and the metadata for all the files and directories in the tree.
In addition, the namenode also knows the datanodes on which all the blocks for a given file are located.
A client accesses the file system on behalf of the user by communicating with the namenode and datanodes.
Datanodes are the laborer of the file system.
They store and retrieve blocks when they are told to (by clients or the namenode).

\subsubsection{How to Read a File}
We now describe the work-flow of reading a file.
It is essential to elucidate this work-flow since it is a major part in the motivation for the work in this chapter
(the work-flow of writing a file is not important to this work).

The client contacts the namenode, giving it the required file's identification, and gets back the locations of the blocks of the file.
For each block, the namenode returns the addresses of the datanodes that have a copy of that block.
Furthermore, the datanodes are sorted according to their proximity to the client (according to the topology of the cluster's network, which is not in the scope of this work).
If the client is itself a datanode (in the case of a \mr~task, for instance), then it will read from the local datanode, if it hosts a copy of the block
(This is called \textit{data locality}, and we will discuss it in more details later).

The client then connects to the first (closest) datanode for the first block in the file, and data is streamed from the datanode back to the client.
When the end of the block is reached, the connection to the datanode will be closed, then find the best datanode for the next block.
The client connects that datanode, and the iteration continues until the end of the file.

The most important aspect of this design is that the client contacts datanodes directly to retrieve data and is guided by the namenode to the best datanode for each block.
This design allows HDFS to scale to a large number of concurrent clients, since the data traffic is spread across all the datanodes in the cluster.
The namenode meanwhile merely has to service block location requests (which it stores in memory, making them very efficient) and does not, for example, serve data, which would quickly become a bottleneck as the number of clients grow.

\subsection{\hd~Job Concepts}
There are two types of nodes in the \hd~cluster that control the job execution process: a \textit{\jtr} and a number of \textit{\ttr}s.

\subsubsection{\jtr}
The \jtr~node is the liaison between the client and the \hd~framework.
Once the client submits code to the cluster, the \jtr~constructs the execution plan by determining which files to process, assigns nodes to different tasks, and monitors all tasks as they're running.
There is only one \jtr~daemon per \hd~cluster.

\subsubsection{\ttr}
The \ttr~is responsible for executing the individual tasks that the \jtr~assigns.
Although there is a single \ttr~per slave node, each \ttr~can spawn multiple OS processes to handle many map tasks or reduce tasks in parallel.
One responsibility of the \ttr~is to constantly communicate with the \jtr.
If the \jtr~fails to receive a heartbeat from a \ttr~within a specified amount of time, it will assume the \ttr~has crashed and will resubmit the corresponding tasks to other nodes in the cluster.

\subsubsection{\hd~Job Work-flow}
In a job run there are four independent entities:

\begin{itemize}
	\item The client, which submits the job.
	\item The \jtr, which coordinates the job run.
	\item The \ttr, which run the tasks that the job has been split into.
	\item The distributed file system (usually HDFS).
\end{itemize}

At start, the client contacts the \jtr~and sends the job information to it.
When the \jtr~receives a call for a job submission, an initialization process begins that involves creating an object to represent the job being run, which encapsulates its tasks, and bookkeeping information to keep track of the tasks' status and progress.
To create the list of tasks to run, the \jtr~first retrieves the input splits computed by the client from the shared file system.
It then creates one map task for each split.
The number of reduce tasks to create is determined by the number of keys created by the map function.

\ttr s run a simple loop that periodically sends heartbeat method calls to the \jtr.
Heartbeats tell the \jtr~that a \ttr~is alive, but they also double as a channel for messages.
As a part of the heartbeat, a \ttr~will indicate whether it is ready to run a new task, and if it is, the \jtr~will allocate it a task, which it communicates to the \ttr~using the heartbeat return value.

To choose a reduce task, the \jtr~simply takes the next in its list of yet-to-be-run reduce tasks, since there are no data locality considerations.
For a map task, however, it takes account of the \ttr's location in the network and selects a task whose input split is as close as possible to the \ttr.
In the optimal case, the task is data-local, that is, running on the same node that the split resides on.
Alternatively, the task may be rack-local: on the same rack, but not the same node, as the split.
Some tasks are neither data-local nor rack-local and retrieve their data from a different rack from the one they are running on, or even from a out-of-network source.

Now that the \ttr~has been assigned a task, the next step is for it to run the task.
Map tasks needs to get some kind of input.
If the input is a file residing on HDFS then the reading process is as described in the previous section.
\hd~supports practically any kind of input (files, SQL queries, REST APIs etc), but the same principle applies to all inputs: unless the data is local to the node running the task, the executing task needs to connect to the data source, open a stream connection, and read the data from it.

When the jobtracker receives a notification that the last task for a job is complete, it simply tells the client that the job has ended.

\subsection{Iterative Algorithms Efficiency Problem} % Maybe think of another name?
\hd~is designed for data processing in a single pass: the user submit a \textit{job} which is composed of map and reduce functions.
When the framework executes that job, the input data is read once during the execution of the map function and is disposed at the end of the job.
\hd~does not directly support the explicit specification of the data repeatedly processed throughout different jobs.
In \hd, the input of a job is always loaded from its original location (be it a distributed file system, a database or any user-specified source).
When applying different jobs on the same data, the repeatedly processed input need to be loaded for each job - creating a considerable disk and network I/O performance overhead.
Many data-mining algorithms, such as clustering and association-rules require iterative computation: the same data are processed again and again until the computation converges or a stopping condition is satisfied.
In \hd~perspective, iterative programs are different jobs executed one-after-the-other.
Therefore, modifying \hd~such that it will support efficient access to the same data in different jobs is a highly-motivated task.

%\subsection{K-Means Clustering in \mr}
%K-means is a classical clustering algorithm that uses an expectation maximization like technique to partition a number of data points into $k$ clusters.
%K-means usually runs on large data sets and can be easily parallelized, which makes it a perfect candidate for \mr~adaptation.
%
%Consider the following \mr~adaptation of the K-means.
%The input for K-means is a (very) large set of points $\db$, a desired number of clusters, $k$ and a termination difference threshold $\epsilon$.
%At first, $k$ initial points from $\db$ are selected as centers.
%Next, distance is calculated (Euclidean distance) from every point in $\db$ to each of the $k$ centers the closest center is selected.
%A messages with the center as key and the point as value emitted by the mapper.
%The reducer collect all of the points of a particular centroid, calculates a new centroid and emits it.
%Once the all the reducers have finished, a new iteration of map and reduce starts, with the new centroids as the selected centers.
%The iterations terminate condition is when difference between old and new centroid is less than or equal to $\epsilon$.

\subsection{Contribution}
In this chapter we present a universal \hd~caching mechanism.
The caching mechanism provides an interface extension to \hd~with the explicit specification of the repeatedly processed input data to be loaded from its source to a local shared memory cache in the executing node. 
The memory is locally shared in the sense that different processes on the same machine can access it.
After the data is loaded into the cache, in each iteration the framework will assign this node the same data, the node will use the data from the cache, reducing significantly the network traffic.

The main contributions of this chapter are:
\begin{itemize}
\item \textit{Universal cache}:
An interface that allows any input data to be cached.
The cache is intended for iterative computations but actually works for every computation on data that is already in the cache.
	
\item \textit{Cache organization model}:
The cached data is organized on a split level, and resides as a key-value pair.
	
\item \textit{Cache management model}:
The \hd~job manager maintains information of the cached data in the nodes, allowing it to assign tasks to nodes already have the needed data in the cache.
	
\item \textit{Java-accessible shared memory}:
A technical solution for sharing memory between different Java virtual machines on the same machine.
	
\item \textit{Implementation}:
We implemented the system by modifying the \hd~framework and adding new classes.
As a result, jobs already developed for \hd~can be used with little modification to use the cache.
	
\item \textit{Experimental evaluation}:
We evaluated the system on iterative jobs that process real-world datasets.
\end{itemize}

%% file: hadoop.cache/tex/cache_related.work.tex
\section{Related Work}
\label{sec:related.work}

Due to (1) the tremendous popularity of \hd, (2) the fact that a great number of applications are already written for it and (3) the vast number of technologies that are built on top of it (like Hive~\cite{hive}, Pig~\cite{pig} and Cascading~\cite{cascading}), a research focused on this framework is needed, and we review only works that are based on it.
Other frameworks such as the prominent Spark~\cite{zaharia2010spark} or Twister~\cite{ekanayake2010twister} are not covered.

Recent papers that address the \iap~of \hd~can be classified into two categories: (1) increasing \textit{data-locality} percentage (when possible) and (2) a caching layer.

\subsection{Data Locality}
Each node in a \hd~cluster may be used both as a processing worker (for executing map or reduce tasks) and as a host for data on the distributed files system.
So, when a new job is scheduled, \hd~tries to run the map-task on the same node that hosts the input data for that task, in order to save the networking cost; if the try is successful, i.e. the node holding the input data is free for processing, than that task is \textit{data-local}.
Data locality rate of a job is the percentage of all data-local tasks in a \mr~job.
A low data locality rate means more data transfer between machines and higher network traffic.

If data locality rates were 100\%, or very close to it, then the \iap~\\
would be less significant.
The lag between iterations would be as low as the rate of reading data from a local disk.
However this is not the case.
Data locality rate may vary: it depends on the data replication factor and the job load on the cluster.
For example, in~\cite{zaharia2010delay} the authors show that a load of 50 jobs can drop data locality rates below 30\%.
However an acceptable rate that takes into account both over and under loading is 70\%.

A trivial way to increase the data locality rate would be to increase the duplication factor of the data, meaning that each data segment will have many copies on different nodes across the cluster, but of course this method requires a storage capacity increase.

A different, more sophisticated, approach, that has been the core issue for many researches, would be to change the scheduling of tasks in the cluster.
Two of them are the most prominent:
Delay Scheduling\cite{zaharia2010delay} is a widely cited paper that presents a simple algorithm that aspire to raise the data locality rate by allowing tasks to wait a limited amount of time before they are executed.
It relaxes the strict job order for task assignment and delays a job execution if the job has no map task local to the current node.
To assign tasks to a node, the delay algorithm starts the task search at the first job in the queue for a local task.
If not successful the scheduler delays the job's execution and searches for a local task from succeeding jobs.
A maximum delay time $D$ is set.
If a job has been skipped longer than $D$, its non-local tasks will then be assigned to execution.
However, this work shows the classic tradeoff between usability and fairness.
No delay means high fairness and no usability, while a long delay means the opposite.
The paper tries to find a golden path, but again, one cannot enjoy both worlds.

Match Making\cite{he2011matchmaking} is another scheduler that considers data locality in its design.
The main idea behind this technique is to give every node a fair chance to grab local tasks before any non-local tasks are assigned to any node.
This paper also relaxes the strict job order for task assignment.
If a local map task cannot be found in the first job, the scheduler will continue searching the succeeding jobs.
In order to give every node a fair chance to grab its local tasks, when a node fails to find a local task in the queue for the first time no non-local task will be assigned to that node.
This means that the node does not receive a map task in this heartbeat interval.
If after the second heartbeat no local task is found, the scheduler will assign a non-local one.
However, when the frequency of job submissions is not high, nodes may turn out to be waiting unnecessarily and finish by not getting data-local tasks. 

%% YGYG I stopped here 15/12/2014
As mentioned, many papers aspire to find a way to increase the data locality rate, however targeting data locality rates has two main flaws that cannot be overcome:
(1) What stands at the base of this approach is the assumption that the data can be localized node-wise.
This assumption is valid in the when a distributed file system is used (in most cases its hdfs).
However, the data can come from a database system or some kind of streaming api and therefore cannot be localized.
(2) 

\subsection{Caching Layer}

HaLoop~\cite{bu2010haloop} is a modified version of \hd~that is designed to serve iterative programs.
HaLoop provides an inter-iteration caching and indexing mechanism on three types of caches: reducer input cache, reducer output cache and mapper input cache.
However, HaLoop caches the data on slave nodes' local disks and not on the slave nodes' RAM, making them load the data from the local disk on every iteration.
In addition, the caching mechanism is applicable only for files that reside on HDFS, and not for the general input format mechanism Hadoop provides.

Dacoop~\cite{liang2011dacoop} is an extended version of Hadoop that introduces a shared memory-based data cache mechanism.
However, like HaLoop, Dacoop supports only caching data that comes from files.

(Our proposed cache mechanism is universal, meaning that is can accept data from any source)

%Spark~\cite{zaharia2012resilient} is an open-source, in-memory data analytics cluster computing framework originally developed by Matei Zaharia in the AMPLab at UC Berkeley.
%In contrast to~\hd 's two-stage disk-based~\mr paradigm, Spark's in-memory primitives provide higher performance for certain applications.
%However, Spark is a different framework from~\hd, meaning that any existing~\hd job would need to be migrated to Spark.

%% file: hadoop.cache/tex/cache_design.tex
\section{Design}
\label{sec:design}
The design of the system is based on the following points:

\subsection{Input Independence}
As mentioned before, the input to a \mr~is usually an HDFS file, but is pluggable, meaning that is can be practically anything.
The cache mechanism can work on any input.

\subsection{RAM based}
The cache will be based on the internal RAM of the executing node.

\subsection{Current \hd~Design}
To understand the design of the modified framework, we first review the current \hd~design.

A \textit{\hd~job} (or simply a \textit{job}) is an execution of one map function followed by a reduce function on a specific data.
A \textit{task} is an execution of map or reduce functions on a specific chunk of the data, called \textit{map-task} or \textit{reduce-task} respectively.
A \textit{node} is a single computer in the cluster.
The \textit{job manager} is a singleton process in the cluster.
It accepts job requests from the users, 

We will now look closely at the life cycle of a \hd~job.

A \hd~job is comprised of map and reduce functions, and a data.
A \hd~job works as follows: the framework assigns a node in the cluster with the given map function and a chunk of the input data (called \textit{input split}).
The framework does its best to ensure that the input split resides on the same computer that executes the map function (i.e. \textit{data-locality} property).
If data-locality cannot be achieved, than the framework aspires to execute the function network-wise as close as possible to the input split: same rack, same segment etc.
At the end of the execution of the function, the node clears its internal RAM, and is ready for the next map function and new data.

Hadoop supports all kind of input data, including databases and various APIs.
However, data-locality is possible only when the input data in a file in the file system.
Moreover, as mentioned earlier, even if the input data is a file in the file system, data-locality is not always guarantied.
This means that Hadoop is designed for large-scale data processing in a single pass.
Many algorithms fit naturally into this model, such as word-counting and sorting.

\subsection{Shared Memory}
\label{subsec:shared.memory}
On Hadoop, a node launches a new Java Virtual Machine (JVM) to run each task in, so that any bugs in the user-defined map and reduce functions don't affect the other tasks (by causing a crash for example).
This means that the local cache for each node must use inter-JVM mechanism.
Our implementation is based on POSIX shared memory~\cite{robbins2003unix}.
Since the mapper accepts the input data to be also key-value pair, the shared memory stores serialized key-value pairs per split.
The shared memory must be synchronized between different tasks because on the first iteration the cache is empty and is loaded by the map-task that is assigned to some data split.
Since more than one map-task can be executed on the same machine, the shared memory must be synchronized.

\subsection{Input Format}
\label{subsec:input.format}
An \textit{input split} is a chunk of the input that is processed by a single map.
Each map processes a single split.
Each split is divided into records, and the map processes each record - a key-value pair - in turn.
Splits and records are logical: there is nothing that requires them to be tied to files, for example, although most of the time, they are.
In a database context, a split might correspond to a range of rows from a table and a record to a row in that range.

\hd~can process practically any type of data format, from flat text files to databases.
However, an interface has to be provided to convert the data format to a logical record.
This interface is called an \textit{input format}.
For example, \hd~is provided with a simple input format for reading text files, where a logical record's value is a single line from the file and the key is the character offset.

In our design we have provided a new input format class implementation called
\linebreak \mintinline{java}{CacheInputFormat}.
This new class uses the Delegation design pattern \cite{deletation_pattern}~to support the pluggable input format feature of \hd.
On the first time a split is read, the input format stores the key-value pairs in the shared memory.
On subsequent use of the same split, the input format fetches the pairs from the shared memory.

The following Java code present the use of the \mintinline{java}{CacheInputFormat}, with a delegated input format for sql queries.
\newline

\begin{minted}{java}
/*
connectionString contains the connection string to the data base.
sqlQuery is the SQL query to execute.
*/
String inputFormatId = connectionString + "|" + sqlQuery;
job.setInputFormatClass(CacheInputFormat.class);
CacheInputFormat.setDelegateInputFormatData(
	job, DBInputFormat.class, inputFormatId
);
\end{minted}

\subsection{Cache Manager}
The cache manager consists of two modules with master-slaves relationship, the master \textit{job-cache-manager},  and the slaves \textit{task-cache-managers}.

\subsubsection{Job Cache Manager}
The job-cache-manager runs on the same machine the \hd~job-manager runs on, and is an integral part of it.
The job-cache-manager orchestrates the cache for all the jobs running on the \hd~cluster.
It keeps track of which worker has which split in the cache, and can send cache-clean commands to tasks in the cluster.
For example, when a node reports to the job-manager that it is free to receive a task, the job-manager tries to assign it with a task such that the input to that task resides in the node's cache.
If the input is not in the cache, then regular task assignment rules holds and the job-cache-manager keeps record of the new cache entry for future iterations or future jobs using the same data.

Its tasks are:
\begin{itemize}
	\item Map splits to nodes.
	\item Prioritizes tasks to nodes that cached the split
	
\end{itemize}

\subsubsection{Task Cache Manager}
The task-cache-manager runs on the nodes in the cluster.
Each \hd~task periodically updates the \hd~job manager about the cache status.
For example, if a node removes a split from the RAM due to no-usage, then the job-manager can instruct other nodes to remove splits from the same file since at least one split will have to be reloaded so it doesn't matter if all the splits will be reloaded.

Its tasks are:
\begin{itemize}
	\item Manages a priority for each cached split.
	\item Affected by cache changes on other nodes.
	\item Notifies the Job Manager upon caching or de-caching a split.
\end{itemize}

%% file: hadoop.cache/tex/cache_experiments.tex
\section{Experiments}
\subsection{Setup}
We have conducted several experiments running the \cfis~algorithm presented in Chapter \ref{ch:mining.closed.itemsets} on the same data but from three different sources, with and without the cache mechanism.
The data is the raw text of tweets from Twitter, and by running the \cfis~algorithm we found word-tuples that are likely to appear together.
The \hd~cluster was composed of 4 Linux machines with 4 GB of memory each.
All machines run \hd~version 2.6.0 with Java 8.
The experiment was performed in the university's campus with a 1 Gbits/s of Internet speed. 

The three data sources are:
\begin{itemize}
	\item 
		\textbf{{Regular HDFS Files}}.
		This is the typical setup of \hd~jobs and will serve as a control group, to make sure the cache mechanism does not dramatically harm the common use case.
		For this setup, we used the Twitter API to download 50 millions tweets (about 5 GB of data) and split and saved them in 50 different files.
		
	\item		
		\textbf{A Remote MySQL relational database}.
		Mainly due to security issues, but also for other reasons, in most commercial network setups the databases resides on a different network segment than the \hd~cluster.
		In this situation data locality is not possible both from the reason of a remote server and from the reason that the data does not reside on HDFS.
		Any access to the data will cost in heavy network traffic, especially iterative access.
		This is a perfect setup to test the cache mechanism.
		
	\item
		\textbf{Twitter API}.
		In most cases where data analysis is to be performed on data from Twitter, the tweets are first harvested and saved to a local storage.
		The reason for this is simple: contacting Twitter is limited by the Twitter API, and also it can create a huge overload on the network.
		Using the cache will make the harvesting procedure superfluous.
\end{itemize}

In the experiments we executed the algorithm on the different datasets, and measured the execution time on the first iteration only and on a full run.

\subsection{Results}

\begin{figure}
	\centering
	\includegraphics [width=10.5cm] {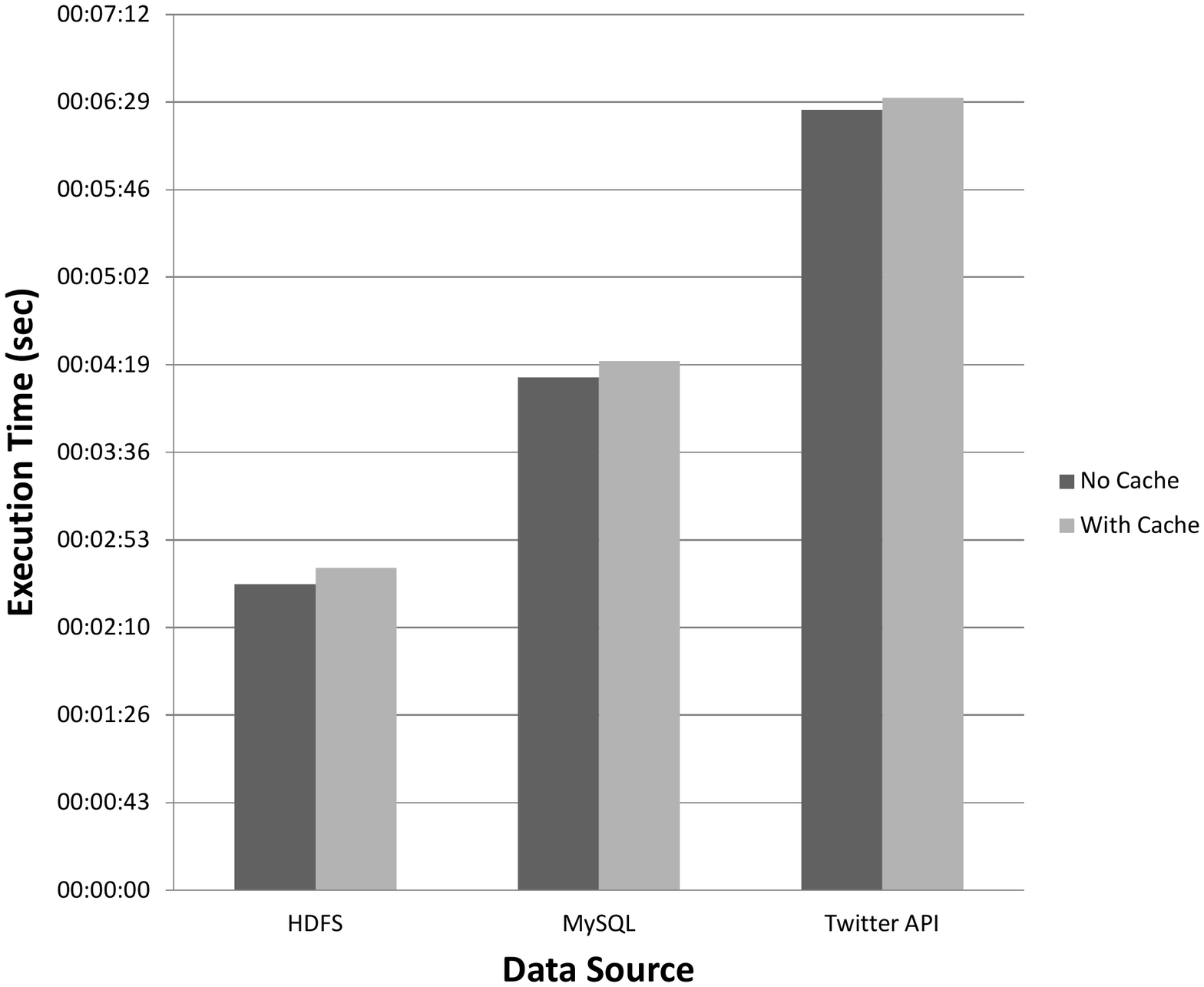}
	\caption{
		Comparing the execution time of the 1\textsuperscript{st} iteration on each of the data sources.
	}
	\label{fig:exp_1st_round}
\end{figure}

\begin{figure}
	\centering
	\includegraphics [width=10.5cm] {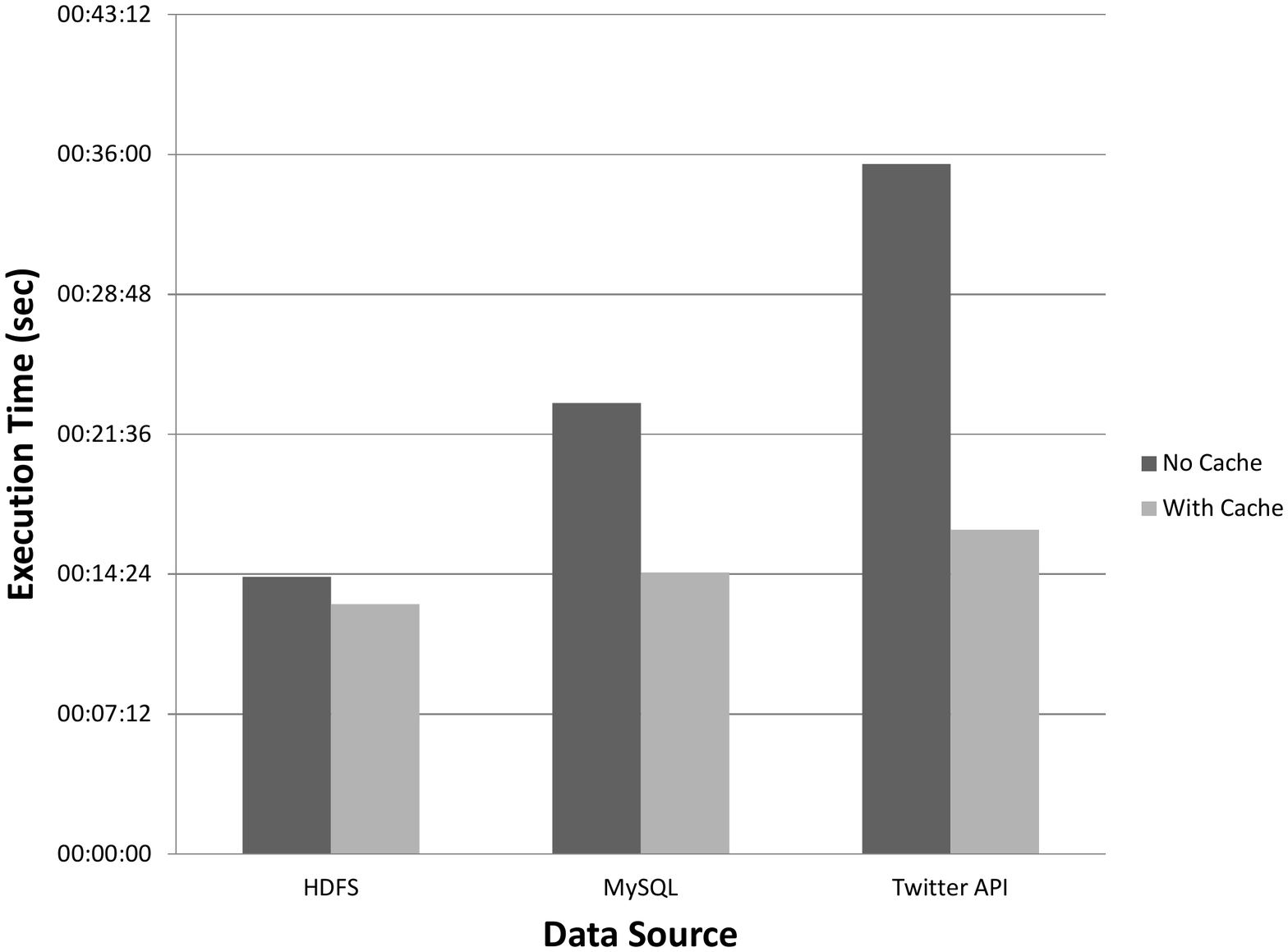}
	\caption{
		Comparing the execution time of the whole job on each of the data sources.
	}
	\label{fig:exp_full_run}
\end{figure}

\begin{figure}
	\centering
	\includegraphics [width=10.5cm] {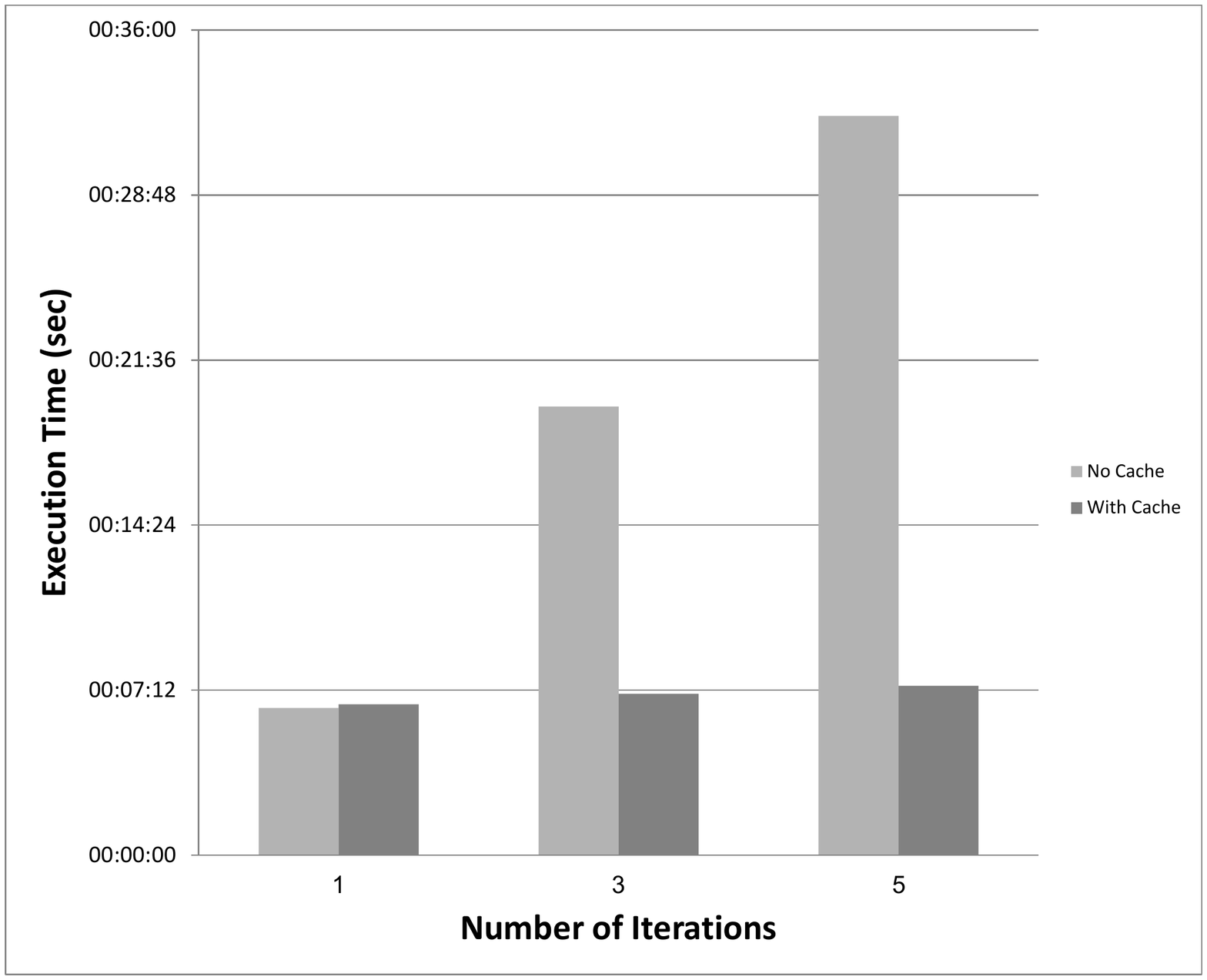}
	\caption{
		Comparing the execution time of different number of iterations on the Twitter API data source.
	}
	\label{fig:exp_twitter}
\end{figure}

At first, we ran the algorithm for mining \cfis~on the three data sources for only a single operation both with and without using the cache mechanism.
The results are shown in Figure~\ref{fig:exp_1st_round}.
As can be seen, with no exceptions, using the cache caused the execution time to be a little slower on all data source.
This was expected because there is some penalty time in loading the split into the cache.

Then we ran the same algorithm on the same three data source but with five iterations.
The results are shown in Figure~\ref{fig:exp_full_run}.
The figure shows that on every data source, using the cache reduces the execution time, especially when using the Twitter API.
When using HDFS files as data source the improvement is not very significant, this is due to two reasons:
\begin{itemize}
	\item 70\% of the map tasks were data local, which means that on 70\% of the input data was not transmitted in the network.
	\item All communications were in the same data segment which is very fast.
\end{itemize}

Lastly, we focused on the Twitter data, and run the \cfis~mining algorithm with increasing number of iteration.
The results are shown in Figure~\ref{fig:exp_twitter}.
The graph shows, as expected, that for the first iteration, the non-cached version is faster than the cached version.
However, for the following iterations, the time addition for the cached algorithm is negligible while the non-cached one is highly less efficient.

%% file: hadoop.cache/tex/cache_conclusion.tex
\section{Conclusion and Discussion}

The experiments undoubtedly demonstrate, that iterative \mr~algorithms, such as the classic K-Means clustering, or mining \cfis, which we used for this work, will highly benefit from a cache mechanism, especially when the data source is not a file residing on HDFS that is deployed in the same network as the \hd~cluster.

%% file: general/future.tex
\chapter{Conclusion and Future Work}
\label{ch:future}

\bookquote{
	Live as if you were to die tomorrow. Learn as if you were to live forever.
}{Mahatma Gandhi}

This dissertation explores the world of processing Big Data.
Our journey began with understanding the challenges data scientists face today when they deal with the deluge of data.
At the top of the list stands the problem that vertical scaling of computer power, i.e. buying a faster computer with more RAM is limited, and the only way to process data and get the results at a reasonable time is to scale horizontally, i.e. working on more computers in parallel.

However, horizontal scaling raises many engineering problems.
The most important ones are:
(i) automating the process of parallelization and distribution of the analysis,
(ii) fault-tolerance, meaning that the analysis of the data will complete even if some of the hardware fails along the way,
(iii) I/O scheduling, meaning that the framework tries to run the task as close as possible to the data, and
(iv) querying the status of the analysis job and monitoring of the process.

In chapter~\ref{ch:map.reduce} we introduced a programming paradigm named \mr~and its most popular implementation, \hd.
\hd~is designed in a way that relieves the programmer of the tedious job of handling the above issues: the programmer needs only to implement two functions: \textit{map} and \textit{reduce}, and the framework takes care of all the rest.

In this dissertation we study the capabilities and limitations of this new tool, and show how it can be used for various tasks that their common trade is that they handle vast amounts of data, and cannot be executed on a single computer.

In the eyes of many programmers and researches \mr~seems like a magic solution for theoretical endless scaling, however some researchers, leaded by Turing prize winner Michael Stonebraker, claim that \mr~is actually a major step backwards \cite{dewitt2008mapreduce}.
The authors have three main claims:
(1) the lack of schemas in \mr~makes applications more buggy and longer to develop,
(2) \mr~does not have any kind of index, like hash or B-tree, and
(3) \mr~is missing important features like updates, transactions and integrity checks.

Since the above paper was published, the big data world was flooded with schema-less NoSQL databases, and other papers improved \mr~with, for example, column based storage \cite{Dittrich:2010:HMY}, so it seems that \mr~does claim its own place among the database world.

In chapter \ref{ch:mining.closed.itemsets} we show that \mr~can be used for complex, non-trivial algorithms like mining \cfis.
Most algorithm published for this mining challenge are not designed for parallel processing and therefore fail to provide a feasible solution for today's data surge.
The algorithm we introduce can scale practically as much as needed, taking advantage of any hardware we can spare.

Chapter \ref{ch:probabilistic} takes us one step further into the world of distributed probabilistic databases, where we build execution plans for queries using \mr.
The challenge here is to develop a query optimizer that does not break the safety of the execution plan.
Big data is by definition not-complete, which makes the probabilistic model a good representation of the real world.
We believe that this is one of the main directions big data research will take.

Finally, we have discovered a weakness in the design of \hd: iterations.
\hd~does not directly support iterations, meaning processing the same data over and over again, with only a small state that changes between iteration.
The workaround used by engineers is creating a job for each iteration.
This workaround is bad because the data needs to be read from its source on each iteration, which makes a huge overload on the network.
The solution we presented is simple and elegant: a cache layer that dictated very small change in how \hd~jobs are coded.

Our journey to the realm of Big Data processing has not yet ended.
Every step of the way, we encountered intriguing questions and alternatives that we constantly continued to explore, and, some areas are still awaiting future research.

%% file: general/apx-free-big-data-sources.tex
\chapter{List of Popular Big Data Repositories}
\label{apx:list.of.big.data.repos}

\begin{itemize}
	\item \textbf{Data.gov} \\
	\url{http://www.data.gov/}\\
	The home of the U.S. Government's open data
	
	\item \textbf{US Census Bureau}\\
	\url{http://www.census.gov/data.html}\\
	The Census Bureau's mission is to serve as a source of data about the nation's people and economy.
	
	\item \textbf{European Union Open Data Portal}\\
	\url{http://data.europa.eu/euodp/en/data/}\\
	The European Union Open Data Portal is a point of access to a growing range of data from the institutions and other bodies of the European Union (EU).
	Data are free to use and reuse for commercial or non-commercial purposes.
	The EU Open Data Portal is managed by the Publications Office of the European Union.
	Implementation of the EU's open data policy is the responsibility of the Directorate-General for Communications Networks, Content and Technology of the European Commission.
	
	\item \textbf{Data.gov.uk}\\
	\url{http://data.gov.uk/}
	
	\item \textbf{The CIA World Factbook}\\
	\url{http://www.cia.gov/library/publications/the-world-factbook/}\\
	The World Factbook provides information on the history, people, government, economy, geography, communications, transportation, military, and transnational issues for 267 world entities.

	\item \textbf{HealthData.gov}
	\url{http://www.healthdata.gov/}
	Making high value health data more accessible to entrepreneurs, researchers, and policy makers in the hopes of better health outcomes for all.
	
	\item \textbf{Amazon Web Services public datasets}\\
	\url{https://aws.amazon.com/datasets/}
	Public Data Sets on AWS provides a centralized repository of public data sets in various topics.
	For Amazon AWS users, this data can be seamlessly integrated into AWS cloud-based applications.

	\item \textbf{Google Books Ngrams}\\
	\url{http://storage.googleapis.com/books/ngrams/books/datasetsv2.html}
	
	\item \textbf{Million Song Data Set}\\
	\url{https://aws.amazon.com/datasets/million-song-dataset/}\\
	The Million Songs Collection is a collection of 28 datasets containing audio features and metadata for a million contemporary popular music tracks.
\end{itemize}